\documentclass[preprint,aip]{revtex4-1}
\usepackage{amsmath,amssymb}
\usepackage{graphicx}
\usepackage{dcolumn}
\usepackage{bm}
\usepackage{color}
\usepackage{rotating}
\usepackage{multirow}

\begin{document}
\title{An analysis of spatiotemporal localized solutions in the variable coefficients (3+1)-dimensional nonlinear Schr\"{o}dinger equation with six different forms of dispersion parameters}
\author{K.~Manikandan}
\affiliation{Centre for Nonlinear Dynamics, Bharathidasan University, Tiruchirappalli - 620 024, Tamilnadu, India}.\\
\author{M.~Senthilvelan}
\email[]{velan@cnld.bdu.ac.in}
\affiliation{Centre for Nonlinear Dynamics, Bharathidasan University, Tiruchirappalli - 620 024, Tamilnadu, India}


\begin{abstract}
We construct spatiotemporal localized envelope solutions of a (3+1)-dimensional nonlinear Schr\"{o}dinger equation with varying coefficients such as dispersion, nonlinearity and gain parameters through similarity transformation technique. The obtained localized rational solutions can serve as prototypes of rogue waves in different branches of science. We investigate the characteristics of constructed localized solutions in detail when it propagates through six different dispersion profiles, namely constant, linear, Gaussian, hyperbolic, logarithm and exponential. We also obtain expressions for the hump and valleys of rogue wave intensity profiles for these six dispersion profiles and study the trajectory of it in each case.  Further, we analyze how the intensity of another localized solution, namely breather, changes when it propagates through the aforementioned six dispersion profiles. Our studies reveal that these localized solutions co-exist with the collapsing solutions which are already found in the (3+1)-dimensional nonlinear Schr\"{o}dinger equation.  The obtained results will help to understand the corresponding localized wave phenomena in related fields.
\end{abstract}
\maketitle
\begin{quotation}
Exploring spatiotemporal localized structures such as solitons, rogue waves (RWs) and breathers are of contemporary interest in several area of physics.  These rational solutions (RWs and breathers) are solutions of nonlinear Schr\"{o}dinger (NLS) equation.  The variable coefficients NLS (vcNLS) equation has been modelled to reflect the inhomogeneities of media, nonuniformities of boundaries, and external forces.  The vcNLS equation helps to control or modify the localized structures, thus shows advantage over systems with constant coefficients. In this work, we consider a (3+1)-dimensional variable coefficients NLS equation and construct the localized rational solutions of it by employing similarity transformation technique.  We investigate the characteristics of constructed localized solutions with six different forms of dispersion profiles.  Further, we analyze the trajectories of first-order rational solutions in the considered dispersion profiles. 
\end{quotation}

\section{Introduction}
During the past ten years or so exploring localized structures in the vcNLS equation \cite{book1,book2,book3} have flourished into a research area of great importance and interest in various contexts \cite{serk,ser,serki,hao,atre:pani}. Unlike the constant coefficient NLS equation, the studies on vcNLS equation reveal that one can control/amplify the localized structures with the help of inhomogeneity parameters \cite{serki,atre:pani,raj:mur,yan,he,kumar,Mani}. Eventhough the one-dimensional equations have given a good understanding on the dynamics, a detailed investigation on the higher-dimensional version (two or three dimensions) of the system will provide a clear visualization about the localized structures.  The vcNLS equation in higher dimensions is not integrable \cite{book3,tower,saito,saka}.  To understand the dynamics of the higher-dimensional systems several investigations have been carried out through numerical simulations \cite{tower,saito,saka,adhi,zyyan,zyan2}.  In fact the studies on higher-dimensional NLS equation have shown that the solutions develop singularities in finite time and blow-up \cite{saka,adhi,zyan2,silber}. Their localized solutions are unstable and subject to collapse in finite time. Since the exact solutions will provide some insights about the considered model, attempts have been made to transform the higher-dimensional vcNLS equation into one-dimensional constant coefficient NLS equation and from the known solutions of the latter equation a family of exact analytical solutions of the former equation has been obtained.  The well-known similarity reduction technique has been invoked to identify the localized structures in vcNLS family of equations \cite{belic,cqdai,zyan,dswang,zjfang,zhu,Mani3}. By employing the same similarity reduction technique a class of solitary wave solutions for the (3+1)-dimensional vcNLS equation have been constructed, see for example Refs. \cite{zjfang,zhu,jiang,lu,schen,petrovic,wang,yywang2}.  Apart from the solitary wave solutions interest has also been shown to obtain localized envelope solutions for the (3+1)-dimensional vcNLS equation \cite{ddfdai,chen,dai2,zheng}.  Such solutions can serve as prototypes of RWs in different branches of science.  The nature of compression/amplification of rational solutions in the (3+1)-dimensional vcNLS equation, in particular RW and breather solutions with different dispersion parameters are yet to be carried out.  In this work, we intend to investigate the variations in intensity profiles of certain localized solutions in a (3+1)-dimensional vcNLS equation with six different dispersion parameters. We note here that the localized envelope solutions which we derive in this work co-exist with the collapsing solutions which are already reported in the considered model. 

RW is a nonlinear wave which appears from nowhere and disappears without a trace \cite{osbrn:rato,osbrn,khar:pelin}. It arises due to the instability of a certain class of initial conditions that tend to grow exponentially and thus have the possibility of increasing up to very high amplitudes. Efforts have been made to explain the RW excitation through a nonlinear process \cite{benj:feir,pere,vishnu,vishnu3}.  It has been found that the rational solutions of the NLS equation can be used to model the RWs \cite{pere,akmv:anki}. Thus the NLS equation helps to describe the structural and dynamical characteristics of this interesting phenomenon. RWs have been first observed in the area of oceanography and then other physical systems including water wave tank experiments \cite{chab}, capillary waves \cite{shatz}, nonlinear optics \cite{solli,kibler} and Bose-Einstein condensates \cite{blud:kono}.  The other localized rational solution that we are interested is breathers.  Physically breathers arise from the effect of modulational instability which is a characteristic feature of various nonlinear dispersive systems and associated with dynamical growth and evolution of periodic perturbation on a continuous background \cite{mande,eleon}. Breathers have been categorized into two main kinds: (i) Akhmediev breather (AB) (periodic in space and localized in time) and (ii) Kuznetsov-Ma solitons (periodic in time and localized in space) \cite{eleon,kuznetsov}.

To analyze the variations in the intensity profiles of the considered localized envelope solutions, we first construct the first- and second-order localized envelope solutions for the (3+1)-dimensional vcNLS equation. We found that the constructed solutions satisfy the (3+1)-dimensional vcNLS equation with a constraint that interrelates the gain/loss and dispersion parameters with nonlinearity parameter.  We then investigate the characteristics of these rational solutions in detail through six different forms of dispersion parameters, namely constant, linear, hyperbolic, exponential, logarithm and Gaussian.  The amplitude of these localized solutions attains the largest in the case of exponential dispersion profile whereas it becomes smaller in the case of logarithmic dispersion profile.  The intensity of these localized solutions increases in all six cases when we increase the value of gain parameter. We then obtain expressions for the hump and valleys of first-order rational solutions.  With the help of these two expressions, we investigate the trajectory of the solution in each one of the cases. Our results reveal that the trajectory of the valleys are not symmetric with respect to propagation distance in all the six dispersion profiles. In addition to the above, we investigate how the intensity of spatially periodic localized wave (AB) changes when it propagates through the aforementioned six different dispersion profiles.   We notice that in the case of constant, linear and Gaussian dispersion profiles the number of peaks in the breather profile decreases with respect to the propagation distance whereas the number of peaks increases/maintains in the cases of hyperbolic, logarithmic and exponential dispersion profiles. 

The paper is organized as follows.  In Sec. II, we construct spatiotemporal localized rational solutions of the (3+1)-dimensional NLS equation with distributed coefficients such as diffraction, nonlinearity and gain parameter.  In Sec. III, we study the characteristics of constructed rational solutions with six different dispersion profiles. In Sec. IV, we obtain expressions for the hump and valleys of RWs and investigate the trajectory of it in all the six dispersion parameters.  In Sec. V, we study the characteristics of breather solution with these six different dispersion profiles. Finally, in Sec. VI, we present our conclusions. 
\section{The Model and its localized solutions}
We consider the following (3+1)-dimensional vcNLS equation \cite{book1,book2,book3}
\begin{equation}
i\psi_z+\frac{\beta(z)}{2}\left(\psi_{xx}+\psi_{yy}+\psi_{tt}\right)+R(z)\vert \psi \vert^2 \psi = i \gamma(z)\psi,
\label{3d:eq1}
\end{equation}
which describes evolution of a slowly varying wave packet envelope $\psi(x,y,t,z)$ in a diffractive nonlinear Kerr medium with anomalous dispersion, in the paraxial approximation.  Here $z$ represents the normalized propagation distance, $x,y$ are the transverse coordinates and $t$ is the retarded time, that is time in the frame of reference moving with the wave packet.  The functions $\beta(z)$, $R(z)$ and $\gamma(z)$ denote the diffraction/dispersion, nonlinearity and gain parameter, respectively and all of them are real analytic function of $z$. In general, Eq. (\ref{3d:eq1}) is not integrable \cite{book3}. We look for an exact analytical solution of (\ref{3d:eq1}) in the form
\begin{equation}
\psi(x,y,t,z)=\rho(z)U(\xi,\tau)\exp[i \phi(x,y,t,z)],
\label{3d:eq2}
\end{equation}
where $\rho(z)$, $\tau(z)$, $\xi(x,y,t,z)$ and $\phi(x,y,t,z)$ are real functions of their arguments and to be determined. The third unknown $U(\xi,\tau)$ is a complex function.

Substituting Eq. (\ref{3d:eq2}) into (\ref{3d:eq1}), we obtain 
\begin{subequations}
\label{3d:eq3}
\begin{eqnarray}
\xi_{xx}+\xi_{yy}+\xi_{tt}=0, \\
\xi_z+\beta(z)(\xi_x\phi_x+\xi_y\phi_y+\xi_t\phi_t)=0, \\
\phi_z+\frac{\beta(z)}{2}(\phi_x^2+\phi_y^2+\phi_t^2)=0, \\
\rho_z(z)+\frac{\beta(z)}{2}\rho(z)(\phi_{xx}+\phi_{yy}+\phi_{tt})-\rho(z)\gamma(z)=0, \\
\tau_z(z)=\rho(z)^2R(z), \;\;\;\;\; \beta(z)(\xi_x^2+\xi_y^2+\xi_t^2)=\rho(z)^2R(z).
\end{eqnarray}
\end{subequations}
Solving these partial differential equations (\ref{3d:eq3}), we get 
\begin{subequations}
\label{3d:eq4}
\begin{align}
\xi(x,y,t,z)&= \frac{(a x + b y+ c t)-\xi_c(z)}{M(z)}, \; \xi_c(z) = -M(z)\int\frac{\beta(z)}{M^2(z)}dz, \\
\rho(z) &= \frac{A_0 \exp (\Gamma (z) \, dz)}{M(z)^{3/2}}, \; \tau(z)=(a^2+b^2+c^2) \int\frac{\beta(z)}{M^2(z)}dz, \\
 M(z)&=m_1+m_0 \int \beta(z) dz, \;\;\;\; \Gamma(z)= \int \gamma(z)dz, \\
\phi(x,y,t,z)&=\frac{M'(z)}{2\beta(z) M(z)}(x^2+y^2+t^2)-\frac{(x+y+t)}{(a+b+c)M(z)} \nonumber \\ &\;\;\;-\frac{3}{2(a+b+c)^2}\int \frac{\beta(z)}{M^2(z)}dz,
\end{align}
\end{subequations}
The parameter $\rho_0$ represents the initial amplitude, $a,b$ and $c$ denote group velocity parameters and $m_0$ and $m_1$ are arbitrary parameters.  The complex function $U(\xi,\tau)$ should satisfy the standard NLS equation, namely 
\begin{equation}
\label{nls}
i \frac{\partial U}{\partial \tau}+\frac{1}{2}\frac{\partial ^2 U}{\partial \xi^2}+ |U|^2 U=0.
\end{equation}
Eq. (\ref{3d:eq2}) be the solution of (\ref{3d:eq1}) provided the following condition is fulfilled, that is 
\begin{eqnarray}
\label{3d:eq5}
R(z)= \frac{\beta (z) M(z) \left(a^2+b^2+c^2\right) e^{-2 \int \gamma (z) \, dz}}{A_0^2}.
\end{eqnarray}
The constraint (\ref{3d:eq5}) interrelates the gain and dispersion parameter with the nonlinearity parameter.  We can choose $\gamma(z)$ and $\beta(z)$ arbitrarily but the nonlinearity parameter $R(z)$ should be fixed as per (\ref{3d:eq5}).

\begin{figure*}[!ht]
\begin{center}
\includegraphics[width=0.9\linewidth]{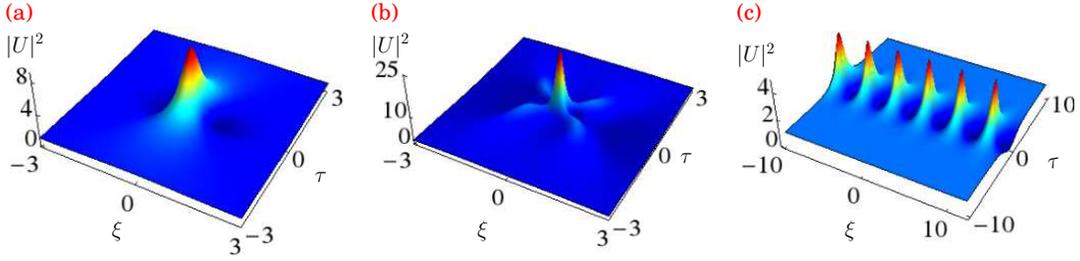}
\end{center}
\caption{(a) First-order RW, (b) Second-order RW and (c) AB of the standard NLS equation.}
\label{nls-rog-br}
\end{figure*}

Eq. (\ref{nls}) admits several localized and periodic solutions \cite{akmv:anki,eleon,kadz}.  As our aim is to construct localized rational solutions such as RW and breather solutions of (\ref{3d:eq1}) we consider the following solutions of (\ref{nls}) \cite{akmv:anki}:
\begin{enumerate}
\item {\bf First-order RW solution} \cite{akmv:anki} \begin{align}
\label{a8}
U_1(\xi,\tau)=\left(1-4\frac{1+2i\tau}{1+4\xi^2+4\tau^2}\right)\exp{[i\tau]}
\end{align}
\item {\bf The second-order RW solution} \cite{akmv:anki}
\begin{subequations}
\label{a11}
\begin{align}
U_2(\xi,\tau)=\left[1+\frac{G_2+i\tau H_2}{D_2}\right]\exp{(i\tau)},
\end{align}
where 
\begin{align}
\label{a9}
G_2 = & \frac{3}{8}-3\xi^2-2\xi^4-9\tau^2-10\tau^4-12\xi^2\tau^2, \notag \\
H_2 = & \frac{15}{4}+6\xi^2-4\xi^4-2\tau^2-4\tau^4-8\xi^2\tau^2,  \\
D_2 = &  \frac{1}{8}\left(\frac{3}{4}+9\xi^2+4\xi^4+\frac{16}{3}\xi^6+33\tau^2+36\tau^4 \right. \notag \\
& \left.+\frac{16}{3}\tau^6-24\xi^2\tau^2+16\xi^4\tau^2+16\xi^2\tau^4\right). \notag
\end{align}
\end{subequations}
\item {\bf Breather solution} \cite{eleon} 
                                                                                                                                                                                                                                                                                                         \begin{align}
U(\xi,\tau) = & \,\biggr[\frac{f^2 \cosh[\alpha (\tau - \tau_1)] + 2 i \,f v \sinh[\alpha (\tau - \tau_1)]}{2  \cosh[\alpha (\tau - \tau_1)] - 2v \cos[ f (\xi - \xi_1) ]}-1\biggr] \exp{(i\tau)},
\label{b1}
\end{align}%
where the parameters $f$ and $v$ are expressed in terms of a complex eigenvalue (say $\lambda$), that is $f=2\sqrt{1+\lambda^2}$ and $v = \mbox{Im}(\lambda)$, and $\xi_1$ and $\tau_1$ serve as coordinate shifts from the origin.  The real part of the eigenvalue represents the angle that the one-dimensionally localized solutions form with the $\tau$ axis, and the imaginary part characterizes frequency of periodic modulation.  The parameter $\alpha$ $(=f v)$ in (\ref{b1}) is the growth rate of modulation instability.
\end{enumerate}
The intensity profiles of these localized structures of the constant coefficient NLS equation are shown in Fig. \ref{nls-rog-br}. 
Incorporating the expressions (\ref{3d:eq4}) into (\ref{3d:eq2}) we can obtain the following solution for (\ref{3d:eq1}), namely 
\begin{eqnarray}
\label{3d:eq6}
\psi(x,y,t,z)&=&\sqrt{a^2+b^2+c^2}\left(\frac{\beta(z)}{R(z)}\right)^{3/2}[U(\xi,\tau)]   \\ & &
\hspace{-3cm} \times \exp \left[{i\left(\frac{M_z(x^2+y^2+t^2)}{2\beta(z) M(z)} -\frac{1}{a+b+c}\frac{(x+y+t)}{M(z)}-\frac{3}{2(a+b+c)^2}\int\frac{\beta(z)}{M^2(z)}dz\right)}-2\int \gamma(z)dz \right].\nonumber
\end{eqnarray}

The obtained localized rational solutions of (\ref{3d:eq6}) can serve as prototypes of RWs and breathers for the model (\ref{3d:eq1}). We also note that such type of rational solutions exist even when random initial conditions are considered at $z$ = 0 \cite{shirara}.  We mention here that the above solutions are constructed from the reduced manifold.  Hence the solutions reflect only 3-fold increase in amplitude and not the blow-up behaviour.  These solutions are special.  The solutions other than these, may form singularities in finite time or blow-up \cite{silber}.  We identify certain novel localized wave structures of (\ref{3d:eq1}) from the obtained solution (\ref{3d:eq6}) by selecting the arbitrary functions appropriately. The arbitrary functions will provide us further choices to develop fruitful structures related to the RWs, which may be useful to raise the feasibility of corresponding experiments and potential applications in physical systems. 
\section{Analysis of RWs in different dispersion parameters}
In this work,  we consider six different dispersion profiles, namely (i) constant, (ii) linear, (iii) Gaussian, (iv) hyperbolic, (v) logarithm and (vi) exponent and investigate the characteristics of the obtained localized solutions under the influence of these six different forms of dispersion profiles. The explicit forms of the considered profiles are \cite{silva,gana,vinoj,amir} 
\begin{subequations}
\label{ed:eq7prof}
\begin{eqnarray}
& (i)\; Constant: \beta(z) &= \frac{1}{\beta_0} \\
&(ii)\;\; Linear:  \beta(z) &= \left(\frac{1-\beta_0}{\beta_0 L}\right) z+ 1 \\
& (iii)\;\; Gaussian: \beta(z) & = \exp\left(\frac{-\ln \beta_0}{L^2} z^2 \right) \\
& (iv)\;\; Hyperbolic: \beta(z) &= \frac{L}{(\beta_0-1)z+L} \\
& (v)\;\; Logarithm: \beta(z) &= \ln\left(e+\frac{z}{L}(e^{(1/\beta_0)}-e)\right) \\
& (vi)\;\; Exponential: \beta(z) &= \exp\left(\frac{-\ln \beta_0}{L} z  \right)
\end{eqnarray}
\end{subequations}
where $\beta_0$ and $L$ are arbitrary parameters.
\begin{figure*}[!ht]
\begin{center}
\includegraphics[width=0.7\linewidth]{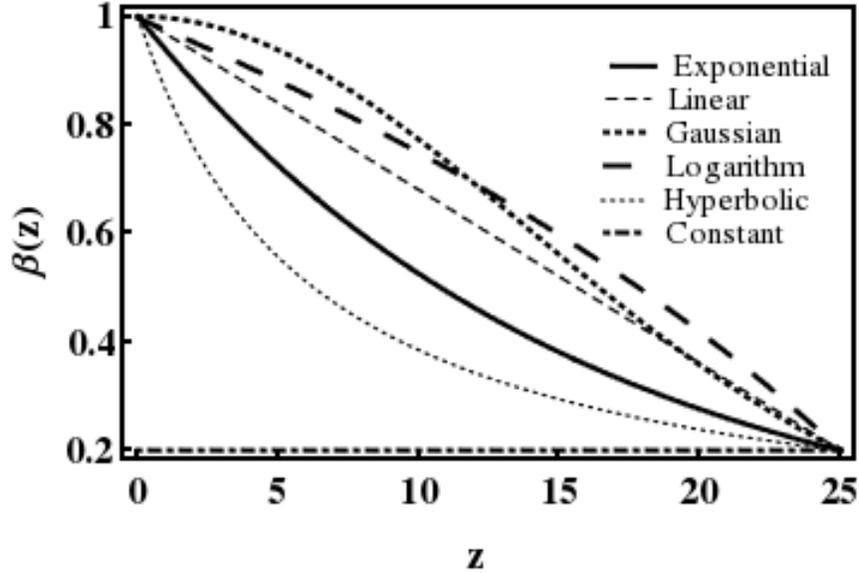}
\end{center}
\caption{Schematic diagrams of different dispersion profiles.  The parameters are fixed as $\beta_0=5$, $L=25$ and $e=2.781$.}
\label{profile}
\end{figure*}
In our analysis, we take $\beta_0=5$ and $L=25$. In Fig. \ref{profile}, we depict the six different dispersion profiles given in (\ref{ed:eq7prof}). We find that the derivative $\frac{d\beta(z)}{dz}$, is continuous and negative for all positive values of $z$, except in the constant dispersion profile \cite{vinoj}.  This in turn confirms that the dispersion decreases monotonically and smoothly except the constant dispersion parameter. The dispersion parameter $\beta(z)$ monotonically decreases from an initial value $1$ to a final value $1/\beta_0$ and it is demonstrated in Fig. \ref{profile}. 

\begin{sidewaystable}
\centering
\caption{The expressions for different forms of $\beta(z), M(z), R(z)$ and $A(z)$.  Here $k_1=a^2+b^2+c^2$, $s_1=\log \left(\frac{z \left(e^{\frac{1}{\beta _0}}-e\right)}{L}+e\right)$ and $s_2=z e^{\frac{1}{\beta _0}}+e (L- z)$.}   
\label{table}
\begin{tabular}{|c|c|c|c|c|}\hline
S.No. &  $\beta(z)$ & $M(z)$ & $R(z)$ & $A(z)$  \\ \hline
1 & $\frac{1}{\beta _0}$ & $\frac{m_0 z}{\beta _0}+m_1$ & $\frac{e^{-2 \gamma _0 t} k_1\left(\beta _0 m_1+m_0 t\right)}{A_0^2 \beta _0^2}$ & $\frac{A_0 e^{\gamma _0 z}}{\left(\frac{m_0 z}{\beta _0}+m_1\right){}^{3/2}}$ \\ \hline
2 & $\frac{\left(1-\beta _0\right) z}{\beta _0 L}+1$ & $\frac{m_0 z \left(\beta _0 (2 L-z)+z\right)}{2 \beta _0 L}+m_1$ & $\frac{e^{-2 \gamma _0 z}k_1 \left(\frac{\left(1-\beta _0\right) z}{\beta _0 L}+1\right) \left(\frac{m_0 z \left(\beta _0 (2 L-z)+z\right)}{2 \beta _0 L}+m_1\right)}{A_0^2}$ & $\frac{A_0 e^{\gamma _0 z}}{\left(\frac{m_0 z \left(\beta _0 (2 L-z)+z\right)}{2 \beta _0 L}+m_1\right){}^{3/2}}$  \\ \hline
3 & $\exp \left(-\frac{z^2 \log \left(\beta _0\right)}{L^2}\right)$ & $\frac{\sqrt{\pi } L m_0 \text{erf}\left(\frac{z \sqrt{\log \left(\beta _0\right)}}{L}\right)}{2 \sqrt{\log \left(\beta _0\right)}}+m_1$ & $\frac{e^{-2 \gamma _0 z} k_1 \beta _0^{-\frac{z^2}{L^2}} \left(\frac{\sqrt{\pi } L m_0 \text{erf}\left(\frac{z \sqrt{\log \left(\beta _0\right)}}{L}\right)}{2 \sqrt{\log \left(\beta _0\right)}}+m_1\right)}{A_0^2}$ & $\frac{A_0 e^{\gamma _0 z}}{\left(\frac{\sqrt{\pi } L m_0 \text{erf}\left(\frac{z \sqrt{\log \left(\beta _0\right)}}{L}\right)}{2 \sqrt{\log \left(\beta _0\right)}}+m_1\right){}^{3/2}}$ \\ \hline
4 & $\frac{L}{L+\left(\beta _0-1\right) z}$ & $\frac{L m_0 \log \left(L+\left(\beta _0-1\right) z\right)}{\beta _0-1}+m_1$ & $\frac{L e^{-2 \gamma _0 z} k_1 \left(\frac{L m_0 \log \left(L+\left(\beta _0-1\right) z\right)}{\beta _0-1}+m_1\right)}{A_0^2 \left(L+\left(\beta _0-1\right) z\right)}$ & $\frac{A_0 e^{\gamma _0 z}}{\left(\frac{L m_0 \log \left(L+\left(\beta _0-1\right) z\right)}{\beta _0-1}+m_1\right){}^{3/2}}$ \\ \hline
5 & $\log \left(\frac{z \left(e^{\frac{1}{\beta _0}}-e\right)}{L}+e\right)$ & $m_0 \left(\frac{e L \log \left(s_2\right)}{e^{\frac{1}{\beta _0}}-e}+\left(s_1-1\right) z\right)+m_1$ & $\frac{s_1 e^{-2 \gamma _0 z} k_1\left(m_0 \left(\frac{e L \log \left(s_2\right)}{e^{\frac{1}{\beta _0}}-e}+\left(s_1-1\right) z\right)\right)}{A_0^2}$ & $\frac{A_0 e^{\gamma _0 z}}{\left(m_0 \left(\frac{e L \log \left(s_2\right)}{e^{\frac{1}{\beta _0}}-e}+\left(s_1-1\right) z\right)+m_1\right){}^{3/2}}$ \\ \hline
6 & $\exp \left(-\frac{z \log \left(\beta _0\right)}{L}\right)$ & $m_1-\frac{L m_0 \beta _0^{-\frac{z}{L}}}{\log \left(\beta _0\right)}$ & $\frac{e^{-2 \gamma _0 z} k_1 \beta _0^{-\frac{z}{L}} \left(m_1-\frac{L m_0 \beta _0^{-\frac{z}{L}}}{\log \left(\beta _0\right)}\right)}{A_0^2}$ & $\frac{A_0 e^{\gamma _0 z}}{\left(m_1-\frac{L m_0 \beta _0^{-\frac{z}{L}}}{\log \left(\beta _0\right)}\right){}^{3/2}}$ \\ \hline  
\end{tabular}
\end{sidewaystable}

\begin{figure*}[!ht]
\begin{center}
\includegraphics[width=0.9\linewidth]{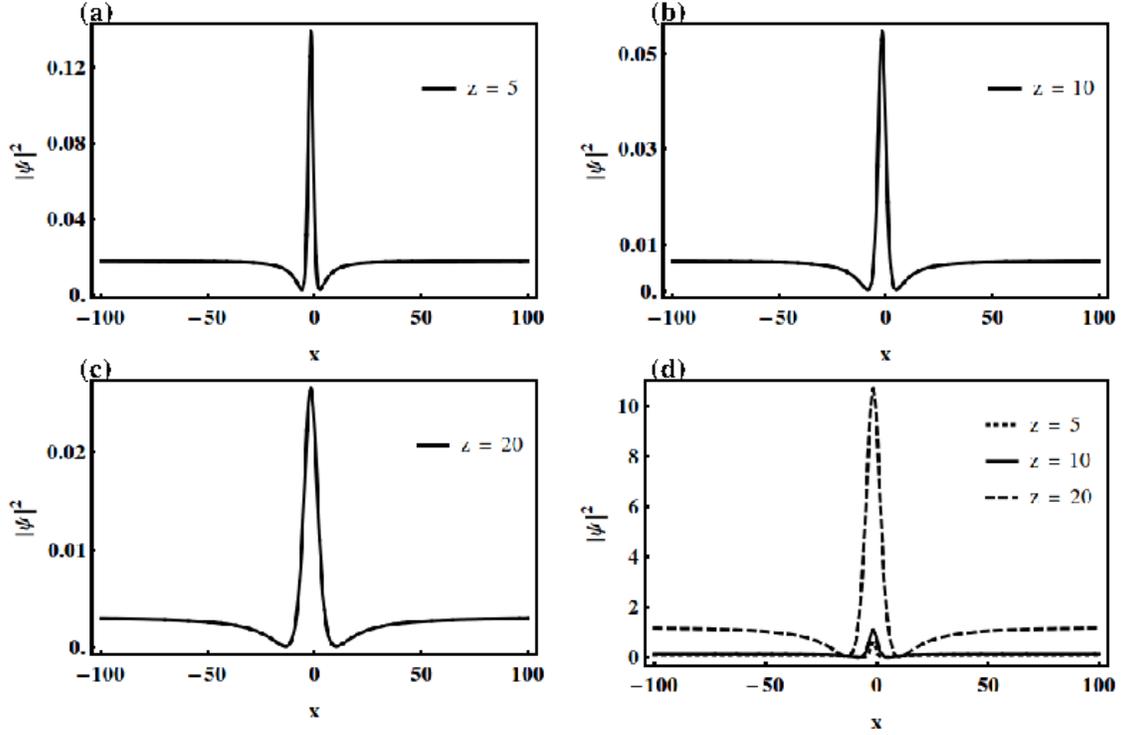}
\end{center}
\caption{The intensity profiles of first-order localized envelope solution for constant dispersion parameter at (a) $z=5$, (b) $z=10$ and (c) $z=20$ with $\gamma_0=0.05$ and (d) $\gamma_0=0.2$. The other parameters are fixed as $y=t=1$, $a=b=c=1$, $m_1=3$, $m_0=0.5$, $\beta_0=5$, $L=25$ and $A_0=1$.}
\label{3d:fig3}
\end{figure*}
\begin{figure*}[!ht]
\begin{center}
\includegraphics[width=0.9\linewidth]{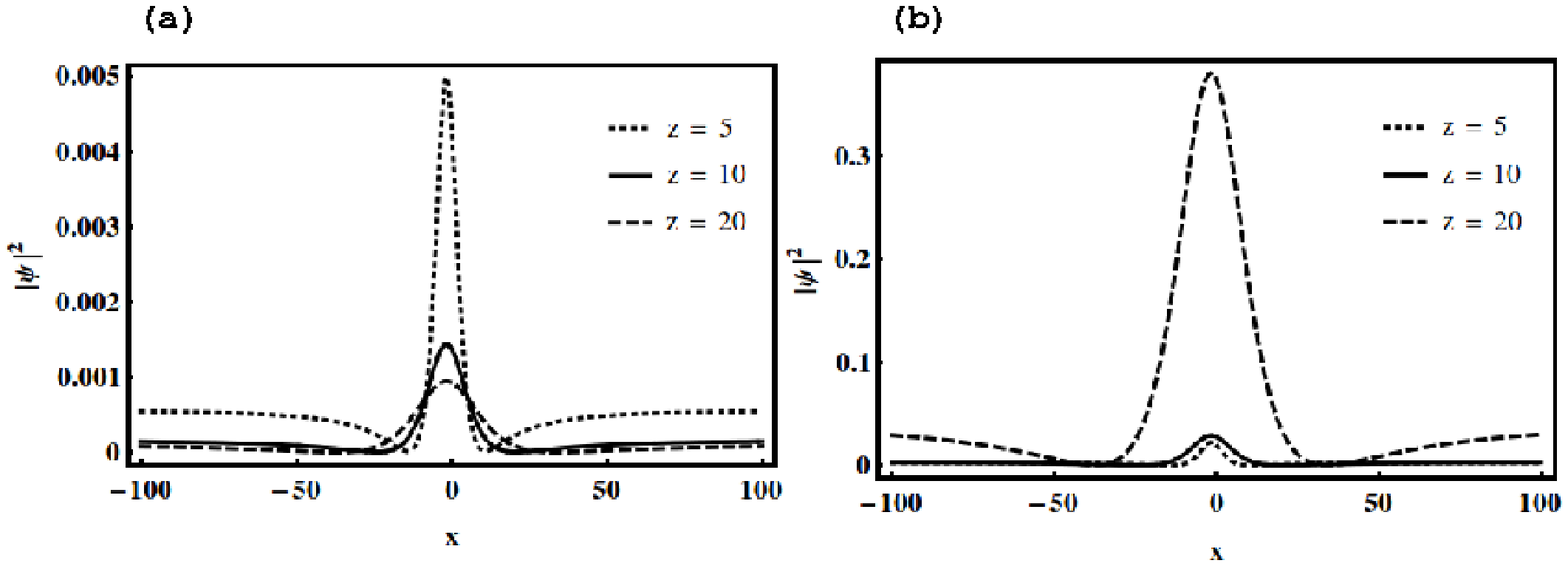}
\end{center}
\caption{The intensity profiles of first-order localized envelope solution for linear dispersion parameter at $z=5, 10$ and $20$ with (a) $\gamma_0=0.05$ and (b) $\gamma_0=0.2$. The other parameters are same as in Fig. \ref{3d:fig3}.}
\label{3d:fig4}
\end{figure*}
\begin{figure*}[!ht]
\begin{center}
\includegraphics[width=0.9\linewidth]{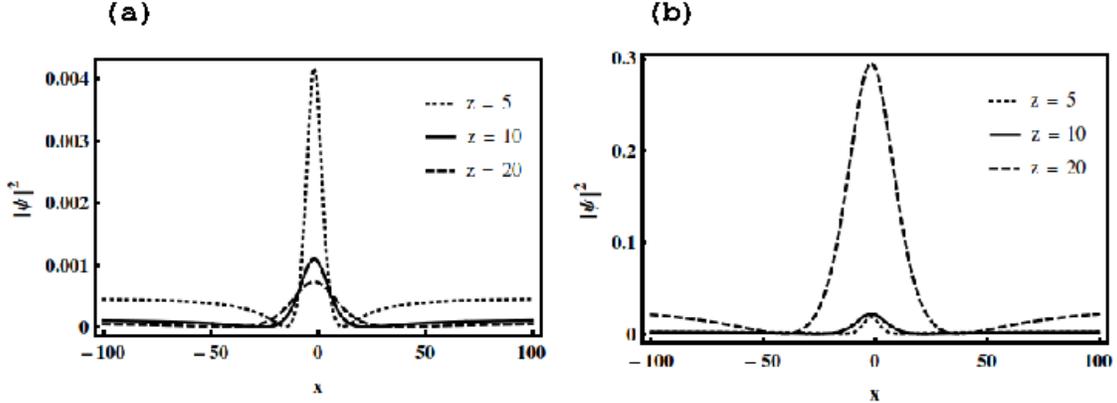}
\end{center}
\caption{The intensity profiles of first-order localized envelope solution for Gaussian dispersion parameter at $z=5, 10$ and $20$ with (a) $\gamma_0=0.05$ and (b) $\gamma_0=0.2$. The other parameters are same as in Fig. \ref{3d:fig3}.}
\label{3d:fig5}
\end{figure*}
\subsection{Characteristics of the first-order localized envelope solution}
For the considered $\beta(z)$, the functions $M(z), R(z)$ and $A(z)$ are found and tabulated in Table \ref{table}. Substituting each one of these expressions separately into (\ref{3d:eq6}) and choosing the rational solution given in (\ref{a8}), we can obtain the first-order localized envelope solution of (\ref{3d:eq1}). This localized envelope solution can serve as a prototypes of first-order RW for the considered model.  In our analysis, we consider the gain parameter $\gamma(z)=\gamma_0$.

Fig. \ref{3d:fig3} shows this first-order rational solution (\ref{3d:eq6}) with constant dispersion parameterfor three different propagation distances, say $z=5$ (Fig. \ref{3d:fig3}(a)), $z=10$ (Fig. \ref{3d:fig3}(b)) and $z=20$ (Fig. \ref{3d:fig3}(c)), when $\gamma_0=0.05$. The figures reveal the usual RW features. At $z=5$, the intensity of this localized profile attains the value of maxima $|\psi|_{max}^2\approx0.15$ as shown in Fig. \ref{3d:fig3}(a).  Intensity of the above localized structure decreases over the propagation distance at $z=10$ and $z=20$ which is demonstrated in Figs. \ref{3d:fig3}(b) and \ref{3d:fig3}(c).  As one observes the structure is more localized at $z=5$ and gradually delocalize at $z=10$ and $z=20$ as shown in Figs. \ref{3d:fig3}(b) and \ref{3d:fig3}(c). The figures further reveal that a decrease in energy in localized profiles.  Normally, the nonlinearity counteracts with dispersion and maintains the shape of the pulse.  In our case the reduced peak power weakens the nonlinearity effect which in turn broadens the localized envelope solution.  To compensate the reduced peak power, we increase the value of gain parameter $(\gamma_0)$ from $0.05$ to $0.2$.  Doing so we observe that the intensity of first-order rational solutions increases as shown in Fig. \ref{3d:fig3}(d). Intensity of this localized profile has now increased as $1$, $1.5$ and $11$ at $z=5$, $10$ and $20$, respectively. The outcome reveals that the intensity of this localized solution increases over the propagation distances.  
\begin{figure*}[!ht]
\begin{center}
\includegraphics[width=0.9\linewidth]{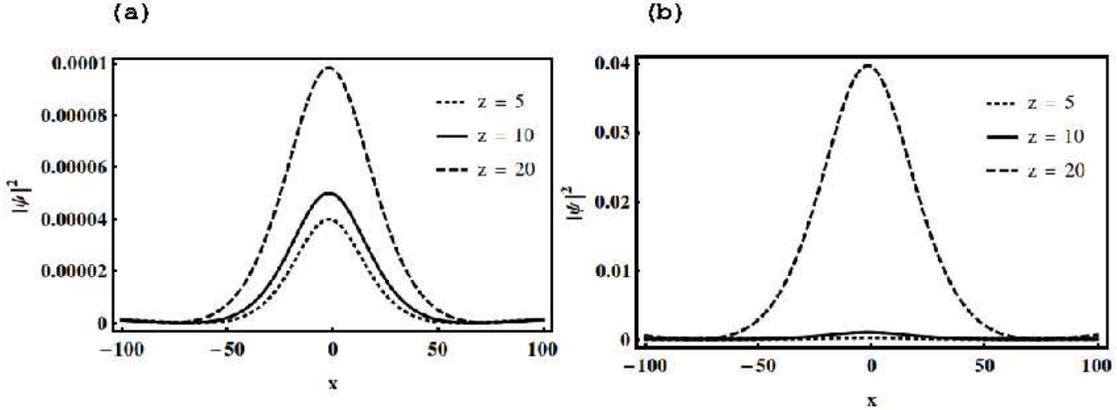}
\end{center}
\caption{The intensity profiles of first-order localized envelope solution for hyperbolic dispersion parameter at $z=5, 10$ and $20$ with (a) $\gamma_0=0.05$ and (b) $\gamma_0=0.2$. The other parameters are same as in Fig. \ref{3d:fig3}.}
\label{3d:fig6}
\end{figure*}
\begin{figure*}[!ht]
\begin{center}
\includegraphics[width=0.9\linewidth]{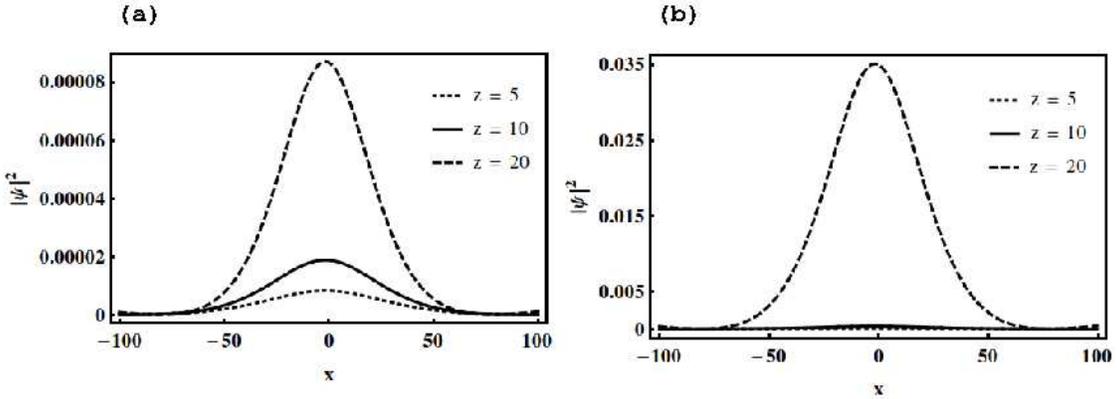}
\end{center}
\caption{The intensity profiles of first-order localized envelope solution for logarithmic dispersion parameter at $z=5, 10$ and $20$ with (a) $\gamma_0=0.05$ and (b) $\gamma_0=0.2$. The other parameters are same as in Fig. \ref{3d:fig3}.}
\label{3d:fig7}
\end{figure*}
Fig.~\ref{3d:fig4} shows the variations in the amplitude of first-order rational solution (optical pulse) described by (\ref{3d:eq6}) for linear dispersion parameter at different distances, that is $z=5, 10$ and $20$, respectively.  The intensity of this localized structure reaches the value of maxima $|\psi|_{max}^2\approx$ $0.005$, $0.0015$ and $0.001$ at $z=5, 10$ and $20$, respectively as shown in Fig. \ref{3d:fig4}(a).  In other words, the intensity of this localized envelope solution attenuates over the propagation distance.  Intensity of the above localized structure increases when we increase the value of $\gamma_0$ from $0.05$ to $0.2$ as shown in Fig.~\ref{3d:fig4}(b).  In Fig.~\ref{3d:fig5}, we display the variations in the amplitude of first-order rational solution for Gaussian dispersion parameter at different propagation distances, that is $z=5, 10$ and $20$.  We notice that the intensity of this localized structure decreases to $0.004 (z=5)$, $0.001 (z=10)$ and $0.0008 (z=20)$, respectively, as shown in Fig.~\ref{3d:fig5}(a). The intensity of the first-order RW is enhanced, for a higher value of $\gamma_0$, say for example $\gamma_0=0.2$, is demonstrated in Fig.~\ref{3d:fig5}(b).

In Fig. \ref{3d:fig6}, we display the first-order localized envelope solution for hyperbolic dispersion profile at different propagation distances, that is $z=5, 10$ and $20$.  The intensity of this localized structure reaches the value of maxima $|\psi|^2\approx0.00004$ at $z=5$ and its intensity increases over propagation distances at $z=10$ and $z=20$ as shown in Fig. \ref{3d:fig6}(a).  We observe that the envelope solution gets more localized at $z=20$ and delocalized at $z=10$ and $z=5$. The intensity of localized structures also increases (increase in energy) which is in contrast with constant, linear and Gaussian dispersion profiles. When we increase the value of $\gamma_0$ from $0.05$ to $0.2$ we observe that intensity of this localized profile increases as illustrated in Fig. \ref{3d:fig6}(b).  The propagation of first-order rational solution through logarithmic dispersion parameter at three different propagation distances, say $z=5, 10$ and $20$, are presented in Fig. \ref{3d:fig7}(a).  At $z=5$, intensity of the above localized structure reaches the value of maxima $|\psi|^2\approx0.000005$ and it increases gradually at $z=10$ and $z=20$ as seen in Fig. \ref{3d:fig7}(a).  We observe that the envelope solution gets more localized at $z=20$ and delocalized at $z=10$ and $z=5$. For a higher value of $\gamma_0$ ($0.2$) the intensity of localized structures gets increased, see Fig. \ref{3d:fig7}(b). 

\begin{figure*}[!ht]
\begin{center}
\includegraphics[width=0.9\linewidth]{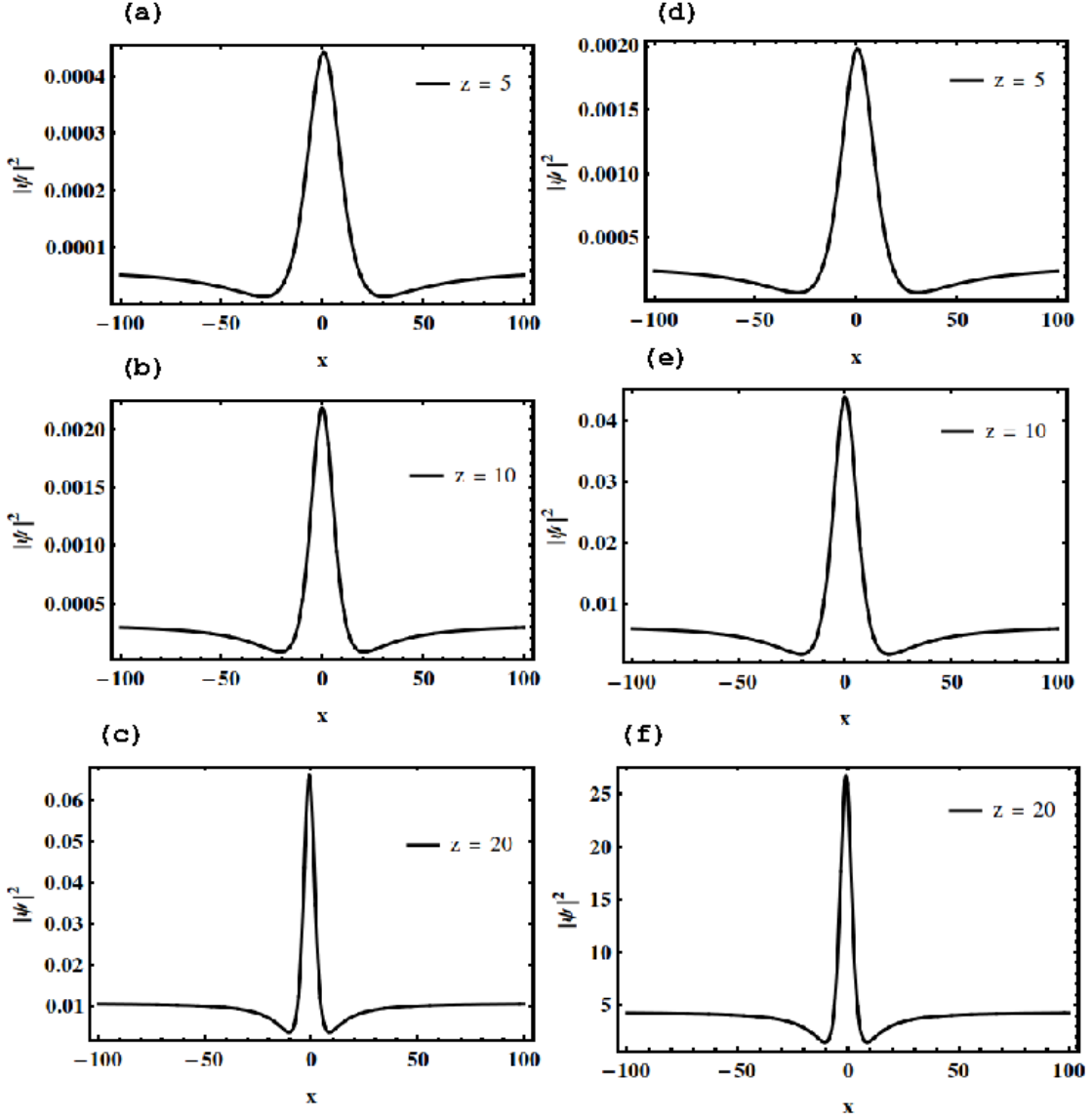}
\end{center}
\caption{The intensity profiles of first-order localized envelope solution for exponential dispersion profile at $z=5, 10$ and $20$ with (a)-(c) $\gamma_0=0.05$ and (d)-(f) $\gamma_0=0.2$. The other parameters are same as in Fig. \ref{3d:fig3} with $m_0=3$ and $m_1=4$.}
\label{3d:fig8}
\end{figure*}

The variations in the amplitude of first-order rational solution for an exponential dispersion profile at different distances, say $z=5, 10$ and $20$, are displayed in Fig. \ref{3d:fig8}. The intensity of this localized structure reaches the value of maxima $|\psi|^2\approx0.0005$ at $z=5$, see Fig. \ref{3d:fig8}(a) and its intensity increases over propagation distances at $z=10$ and $z=20$ as shown in Figs. \ref{3d:fig8}(b) and (c), respectively.  The intensity of envelope solution is enhanced, for $\gamma_0=0.2$, and is demonstrated in Figs. \ref{3d:fig8}(d)-(f).  Differing from the earlier cases, in this profile, we observe the compression of first-order localized envelope solution. 

From the outcome, we conclude that the same initial pulse (first-order rational solution) when propagates through different forms of dispersion profiles has attain different amplitudes and widths.  Among the six dispersion profiles, we have considered the amplitude of the RW in the exponential dispersion profile attains the largest whereas the amplitude in the logarithmic dispersion profile turns out to be the smallest.  
\begin{table}[]
\centering
\caption{The effect of different dispersion profiles in the intensity of localized envelope solutions.}
\label{int-tab}
\begin{tabular}{|c|c|c|c|}
\hline
\multirow{2}{*}{} & \multirow{2}{*}{} & \multicolumn{2}{c|}{Intensity of} \\ \cline{3-4} 
S.No. & Dispersion profile  & First-order envelope solution  & Second-order envelope solution \\ \hline
1 & constant & decreases & decreases  \\ \hline
2 & linear & decreases & decreases  \\ \hline
3 & Gaussian & decreases & decreases  \\ \hline
4 & hyperbolic & increases & increases  \\ \hline
5 & logarithm & increases & increases  \\ \hline
6 & exponent & increases,compressed & increases,compressed \\ \hline 
\end{tabular}
\end{table}
\subsection{Characteristics of the second-order localized envelope solution}
\begin{figure*}[!ht]
\begin{center}
\includegraphics[width=0.9\linewidth]{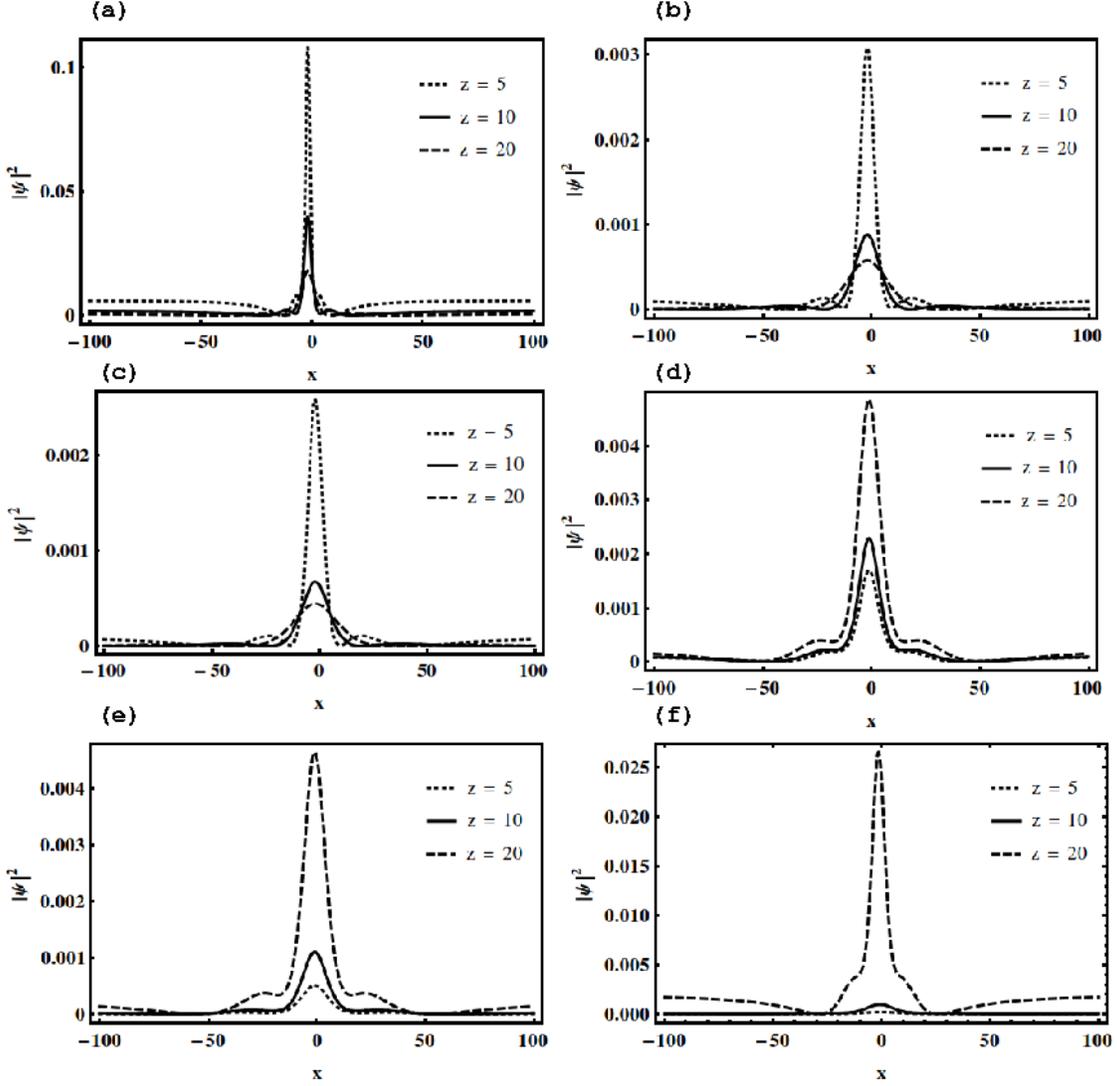}
\end{center}
\caption{The intensity profiles of second-order localized envelope solution at $z=5, 10$ and $20$ for (a) constant, (b) linear, (c) Gaussian, (d) hyperbolic, (d) logarithmic and (e) exponential dispersion profile with $\gamma_0=0.05$. The other parameters are same as in Fig. \ref{3d:fig3}.}
\label{3d:fig9}
\end{figure*}

We can obtain the second-order localized envelope solution of (\ref{3d:eq1}) by substituting the rational solution given in (\ref{a11}) and the chosen $\beta(z)$ in (\ref{3d:eq6}). This localized envelope solution can serve as prototypes of second-order RW solution for the considered model.  Fig. \ref{3d:fig9} represents the intensity profiles of the second-order rational solution for six different forms of dispersion profiles when $\gamma_0=0.05$.  The sequence of propagation of second-order localized envelope solution through constant, linear, Gaussian, hyperbolic, logarithmic and exponential dispersion profiles at different propagation distances, say $z=5$, $z=10$ and $z=20$ are depicted in this figure. The second-order rational solution also reveals the same characteristics, as that of first-order rational solution, when it propagates through the dispersion parameters. 

\begin{figure*}[!ht]
\begin{center}
\includegraphics[width=0.9\linewidth]{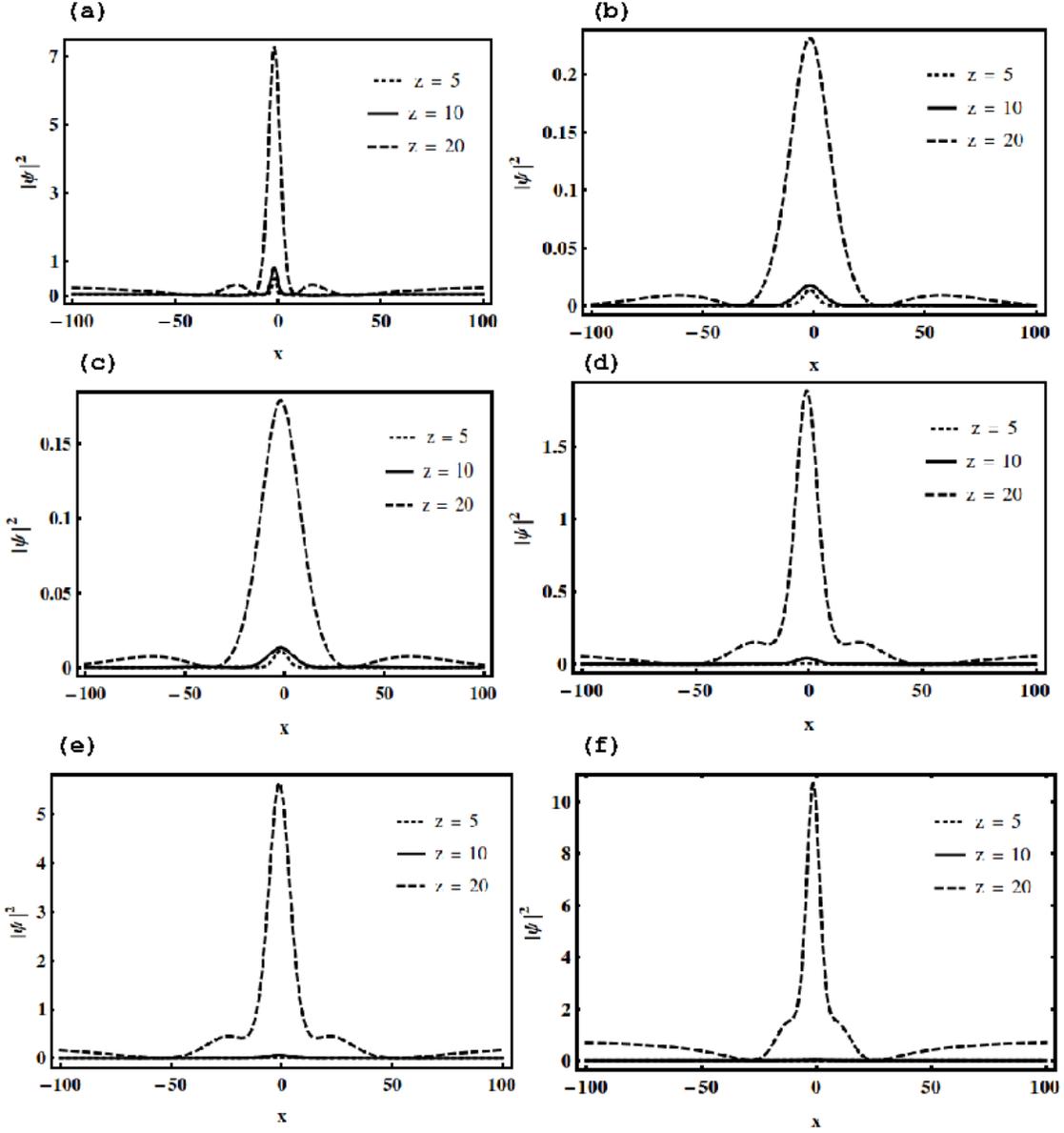}
\end{center}
\caption{The intensity profiles of second-order localized envelope solution at $z=5, 10$ and $20$ for (a) constant, (b) linear, (c) Gaussian, (d) hyperbolic, (d) logarithmic and (e) exponential dispersion profile with $\gamma_0=0.2$. The other parameters are same as in Fig. \ref{3d:fig3}.}
\label{3d:fig10}
\end{figure*}

In the case of constant (Fig. \ref{3d:fig9}(a)), linear (Fig. \ref{3d:fig9}(b)) and Gaussian (Fig. \ref{3d:fig9}(c)) profiles the amplitudes of this second-order localized envelope solution attenuates over the propagation distances whereas its amplitude increases in the case of hyperbolic (Fig. \ref{3d:fig9}(d)), logarithmic (Fig. \ref{3d:fig9}(e)) and exponential (Fig. \ref{3d:fig9}(f)) dispersion profiles. We notice that the above localized structure gets compressed in the case of exponent dispersion profile.  The amplitude of second-order localized envelope solution in the exponential profile becomes the largest and the amplitude in the logarithmic profile becomes the smallest.  When we increase the value of the gain parameter $\gamma_0$ from $0.05$ to $0.2$, to compensate the reduced peak power, the intensity of the second-order rational solution gets increased in all the profiles as shown in Figs. \ref{3d:fig10}(a)-(f). 

In Table \ref{int-tab}, we summarize the variations in the intensity of first-order and second-order envelope solutions for the six different forms of dispersion parameters which we have considered.   
\section{Trajectories of first-order localized envelope solution}
In this section, we investigate the following characteristics of first-order localized envelope solution, namely (i) the evolution of its hump, (ii) its width (distance between the two valleys) and (iii) the nature of its trajectory, analytically.  The trajectory of this type of localized structure is usually described by the motion of its hump and the valleys \cite{ling}.  Since we have an exact expression for the solution, we can calculate the position of peak and the two valleys of this profile.  For this purpose, we apply the extremum theorem to Eq. (\ref{3d:eq6}) and fix $y=t=1$, namely $[\partial \vert \psi(x,y,t,z)\vert^2/\partial x]_{x=x_c}=0$, and obtain
\begin{align}
\label{cubic}
\left(a m_0 e^{\gamma _0 z} \left(a^2+b^2+c^2\right) \left(m_0 (a x_c+b+c)-1\right) \left(4 \left(3 \left(a^2+b^2+c^2\right)^2-1\right) \right. \right. \nonumber \\ \left. \left. +m_ 0\left (3\left (m_ 0 (\int\beta (z)\, dz) + m_ 1 \right){}^2 - 4 (a x_c + b + c)^2 \right) \right) \right) =0.
\end{align}
From (\ref{cubic}), we can obtain
\begin{subequations}
\label{traj}
\begin{eqnarray}
\label{hump} a m_0 e^{\gamma _0 z} \left(a^2+b^2+c^2\right) \left(m_0 (a x_h+b+c)-1\right)&=&0 , \\
\label{valley} m_0 \left(m_0 \left(3 \left(m_0 (\int \beta (z) \, dz)+m_1\right){}^2-4 (a x_v+b+c)^2\right)+8 (a x_v+b+c)\right) \nonumber \\
+ 4 \left(3 \left(a^2+b^2+c^2\right)^2-1\right)&=&0.
\end{eqnarray}
\end{subequations}

\begin{figure}[!ht]
\begin{center}
\includegraphics[width=0.9\linewidth]{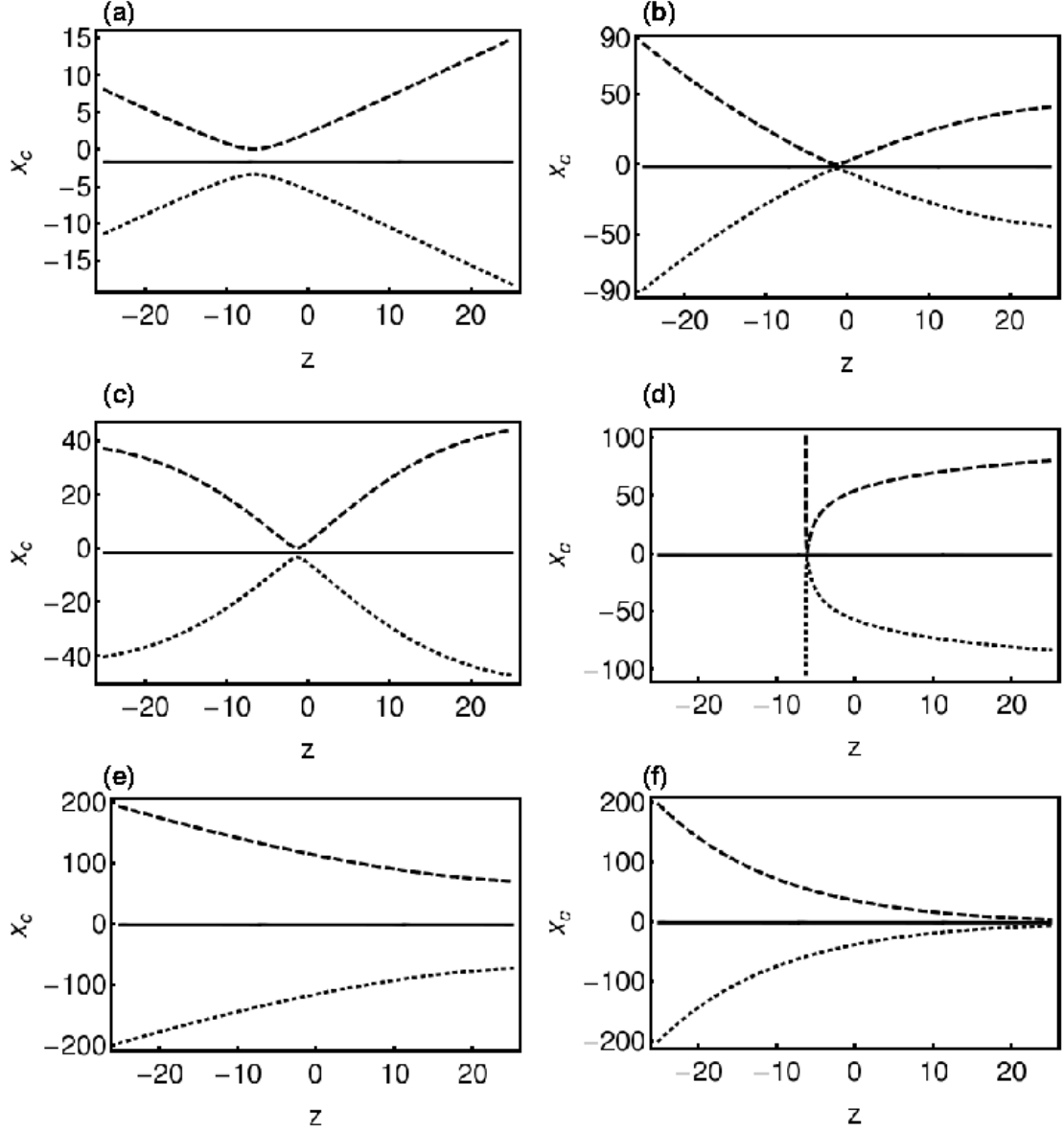}
\end{center}
\caption{The trajectory of localized envelope solution (\ref{3d:eq6}) for (a) $\beta(z) = \frac{1}{\beta_0}$, (b) $\beta(z)=\left(\frac{1-\beta_0}{\beta_0 L}\right) z+ 1$, (c) $\beta(z) = \exp\left(\frac{-z^2}{L^2} \ln \beta_0 \right)$, (d) $\beta(z) = \frac{L}{(\beta_0-1)z+L}$, (e) $\beta(z) = \ln\left(e+\frac{z}{L}(e^{(1/\beta_0)}-e)\right)$ and (f) $\beta(z) = \exp\left(\frac{-z}{L} \ln \beta_0 \right)$.  The vertical axes $x_c$ denotes $x_h$, $x_{v_1}$ and $x_{v_2}$.  The solid line represents the hump of RW $x_h$, dashed and dotted line represent the motions of the two valleys $x_{v_1}$ and $x_{v_2}$. The other parameters are same as in Fig. \ref{3d:fig3}.} 
\label{fig8}
\end{figure}

By solving Eq. (\ref{traj}), we can find the position of hump and the two valleys ($x_h$, $x_{v_1}$ and $x_{v_2}$) with respect to propagation distance $(z)$.  From Eq. (\ref{hump}) we can obtain the expression for position of hump ($x_h$), that is 
\begin{equation}
\label{xhump}
x_h = \frac{1-m_0 (b+c)}{a m_0}.
\end{equation}
The second expression (\ref{valley}) yields the expression for positions of the two valleys ($x_{v_1}$, $x_{v_2}$), that is 
\begin{align}
\label{xval}
 x_{v_1,v_2}= \frac{2 a m_0 \left(1-m_0 (b+c)\right) \pm \sqrt{3} \sqrt{a^2 m_0^2 \left(4 \left(k_1\right)^2+m_0^2 \left(m_0 (\int \beta (z) \, dz)+m_1\right){}^2\right)}}{2 a^2 m_0^2},
\end{align}
where $k_1=a^2+b^2+c^2$.
\begin{sidewaystable}
\centering
\caption{The expressions of the hump and the two valleys of RW intensity profiles for different forms of $\beta(z)$.}.  
\label{table2}
\begin{tabular}{|c|c|c|c|}\hline
S.No. &  $\beta(z)$ & $x_h$ & $x_{v_1,v_2}$   \\ \hline
1 & $\frac{1}{\beta_0}$  & $\frac{1-m_0 (b+c)}{a m_0}$ & $\frac{\pm \sqrt{3} \sqrt{a^2 m_0^2 \left(4 k_1^2+m_0^2 \left(\frac{m_0 z}{\beta _0}+m_1\right){}^2\right)}+2 a m_0 \left(1-m_0 (b+c)\right)}{2 a^2 m_0^2}$  \\ \hline
2 & $\left(\frac{1-\beta_0}{\beta_0 L}\right) z + 1$  & $\frac{1-m_0 (b+c)}{a m_0}$ &  $ \frac{\pm \sqrt{3} \sqrt{a^2 m_0^2 \left(4 k_1^2+m_0^2 \left(\frac{m_0 z \left(\beta _0 (2 L-z)+z\right)}{2 \beta _0 L}+m_1\right){}^2\right)}+2 a m_0 \left(1-m_0 (b+c)\right)}{2 a^2 m_0^2}$  \\ \hline
3 & $\exp\left(\frac{-z^2}{L^2} \ln \beta_0 \right)$  & $\frac{1-m_0 (b+c)}{a m_0}$ & $ \frac{\pm \sqrt{3} \sqrt{a^2 m_0^2 \left(m_0^2 \left(\frac{\sqrt{\pi } L m_0 \text{erf}\left(\frac{z \sqrt{\log \left(\beta _0\right)}}{L}\right)}{2 \sqrt{\log \left(\beta _0\right)}}+m_1\right){}^2+4 k_1^2\right)}+2 a m_0 \left(1-m_0 (b+c)\right)}{2 a^2 m_0^2}$  \\ \hline
4 & $\frac{L}{(\beta_0-1)z+L}$  & $\frac{1-m_0 (b+c)}{a m_0}$ & $ \frac{\pm \sqrt{3} \sqrt{a^2 m_0^2 \left(4 k_1^2+m_0^2 \left(\frac{L m_0 \log \left(L+\left(\beta _0-1\right) z\right)}{\beta _0-1}+m_1\right){}^2\right)}+2 a m_0 \left(1-m_0 (b+c)\right)}{2 a^2 m_0^2}$  \\ \hline
5 & $\ln\left(e+\frac{z}{L}(e^{(1/\beta_0)}-e)\right)$  & $\frac{1-m_0 (b+c)}{a m_0}$ & $ \frac{\pm \sqrt{3} \sqrt{a^2 m_0^2 \left(m_0^2 \left(m_0 \left(z \left(\log \left(\frac{z \left(e^{\frac{1}{\beta _0}}-e\right)}{L}+e\right)-1\right)+\frac{e L \log \left(z e^{\frac{1}{\beta _0}}+e L-e z\right)}{e^{\frac{1}{\beta _0}}-e}\right)+m_1\right){}^2+4 k_1^2\right)}+2 a m_0 \left(1-m_0 (b+c)\right)}{2 a^2 m_0^2}$  \\ \hline
6 & $\exp\left(\frac{-z}{L} \ln \beta_0 \right)$  & $\frac{1-m_0 (b+c)}{a m_0}$ & $ \frac{\pm \sqrt{3} \sqrt{a^2 m_0^2 \left(4 k_1^2+m_0^2 \left(m_1-\frac{L m_0 \beta _0^{-\frac{z}{L}}}{\log \left(\beta _0\right)}\right){}^2\right)}+2 a m_0 \left(1-m_0 (b+c)\right)}{2 a^2 m_0^2}$   \\ \hline
\end{tabular}
\end{sidewaystable}
The exact expressions of the hump and valleys of the first-order rational solution for different $\beta(z)$ are tabulated (see Table \ref{table2}) and plotted in Fig. {\ref{fig8}}.  
The maximum intensity of this localized profile is found to be 
\begin{eqnarray}
|\psi_1|^2_{max}=\frac{k_1 e^{2 \gamma _0 z} \left(4 k_1^2+9 m_0^2 \left(m_0 (\int \beta (z) \, dz)+m_1\right){}^2\right)}{\left(m_0 (\int \beta (z) \, dz)+m_1\right){}^3 \left(4 k_1^2+m_0^2 \left(m_0 (\int \beta (z) \, dz)+m_1\right){}^2\right)}
\end{eqnarray}
and the minimum intensity is found to be
\begin{eqnarray}
|\psi_1|^2_{min}=\frac{4 k_1^3 e^{2 \gamma _0 z}}{\left(m_0 (\int \beta (z) \, dz)+m_1\right){}^3 \left(4 k_1^2+m_0^2 \left(m_0 (\int \beta (z) \, dz)+m_1\right){}^2\right)}.
\end{eqnarray}
The width of the envelope solution (distance between two valleys), which evolves with propagation distance, is given by
\begin{eqnarray}
W(t)=\frac{\sqrt{3 a^2 m_0^2 \left(4 k_1^2+m_0^2 \left(m_0 (\int \beta (z) \, dz)+m_1\right){}^2\right)}}{a^2 m_0^2}.
\end{eqnarray}

Using the above expressions now we analyze the evolution of the hump and valleys of the RW intensity profiles.  Fig. \ref{fig8} represents the trajectories of localized structure for different dispersion profiles $(\beta(z))$ with $\beta_0=5$ and $L=25$.  In this figure, the solid line denotes the trajectory of the hump whereas the dotted and dashed lines indicate the motion of two valleys of the localized solution $(x_h)$. We begin our analysis by considering the constant dispersion parameter $\beta(z) = \frac{1}{\beta_0}$. The corresponding trajectory of the first-order rational solution is depicted in Fig. \ref{fig8}(a).  It is clear from this figure that the trajectory of the extrema of the envelope solution travels almost in a straight line and the two valleys come closer to each other at $z=6$. We observe that $|\psi|^2_{max}$ travels in a straight line in the neighbourhood of the origin whereas the valleys travel in an $'X'$ shaped path which in turn reveals the fundamental RW feature. Fig. \ref{fig8}(b) shows the trajectory of localized structure for $\beta(z)=\left(\frac{1-\beta_0}{\beta_0 \, L}\right)z + 1$.  At $z=-20$, the positions of the two valleys are well separated. The two valleys come closer to the hump position near $z=0$.  The position of the valleys depart from each other soon after as shown in Fig. \ref{fig8}(b).  In Fig. \ref{fig8}(c), we depict the trajectory of obtained envelope solution for $\beta(z) = \exp\left(\frac{-z^2}{L^2} \ln \beta_0 \right)$.  Here we observe that the position of the two valleys are well separated at $z=-20$ and approach one another slowly and give rise to the localized structure near $z=0$.  Soon after, we find that the trajectories rapidly deviate from each other.  The trajectory of the envelope solution for the hyperbolic type dispersion profile $\beta(z) = \frac{L}{(\beta_0-1)z+L}$ is shown in Fig. \ref{fig8}(d). The motion of the two valleys meet each other at $z\approx-7$ and then they deviate from each other slowly. Trajectory of the first-order rational solution of $\beta(z) = \ln\left(e+\frac{z}{L}(e^{(1/\beta_0)}-e)\right)$ is shown in Fig. \ref{fig8}(e). In this figure, we notice that the motion of the two valleys travels in a decreasing manner. Differing from the earlier four cases here the motion of the two valleys decreases monotonically. In Fig. \ref{fig8}(f), we present the trajectory of the considered first-order rational solution for $\beta(z) = \exp\left(\frac{-z}{L} \ln \beta_0 \right)$.  Here we notice that the position of two valleys decreases very rapidly from $z=-20$ to $z\approx 20$.  We observe that the trajectories of the two valleys are not symmetric with respect to propagation distance in any of the forms of dispersion profiles considered in this work.   
\section{Analysis of breather in different forms of dispersion parameters}
In the above, we have analyzed how the intensity profiles of RWs get modified by the dispersion profiles in the (3+1)-dimensional vcNLS equation (\ref{3d:eq1}).  Now, we investigate the dynamics of other rational solution, namely, breather with respect to different forms of dispersion parameters.  To study the breather dynamics of (\ref{3d:eq1}), we substitute the breather solution (\ref{b1}) of NLS equation into (\ref{3d:eq6}).  We can obtain the AB solution, which is periodic in space and localized in time, for (\ref{3d:eq1}) by choosing the eigenvalue $\lambda=0.65 i$ in (\ref{b1}). 
\begin{figure*}[!ht]
\begin{center}
\includegraphics[width=0.9\linewidth]{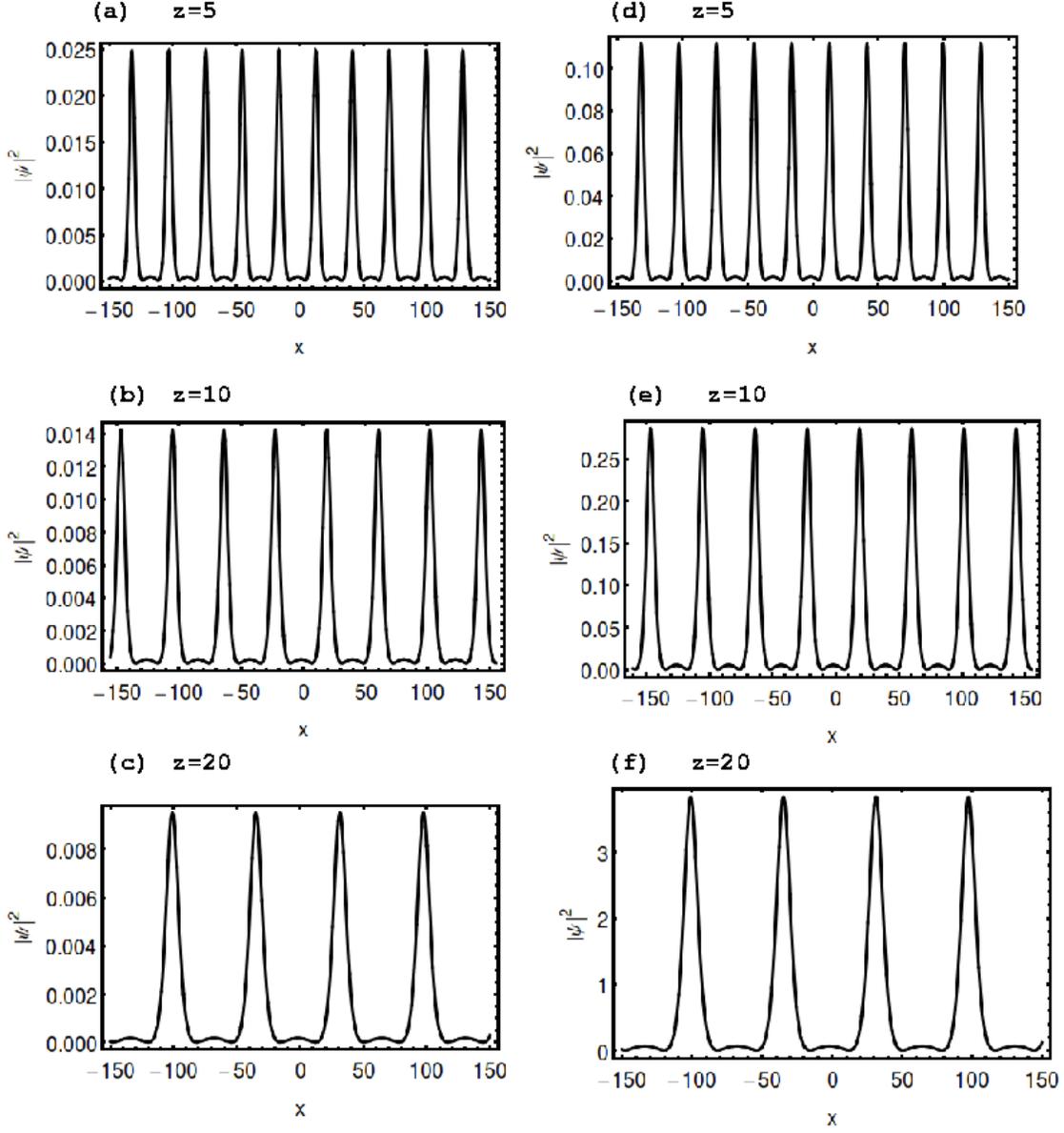}
\end{center}
\caption{The intensity profiles of AB for constant dispersion parameter at (a) $z=5$, (b) $z=10$ and (c) $z=20$ with $\gamma_0=0.05$ and (d) $z=5$, (e) $z=10$ and (f) $z=20$ with $\gamma_0=0.2$. The other parameters are fixed as $y=t=1$, $a=b=c=1$, $m_1=3$, $m_0=4$, $\beta_0=5$, $L=25$ and $A_0=1$.}
\label{3d:fig12}
\end{figure*}
\begin{figure*}[!ht]
\begin{center}
\includegraphics[width=0.9\linewidth]{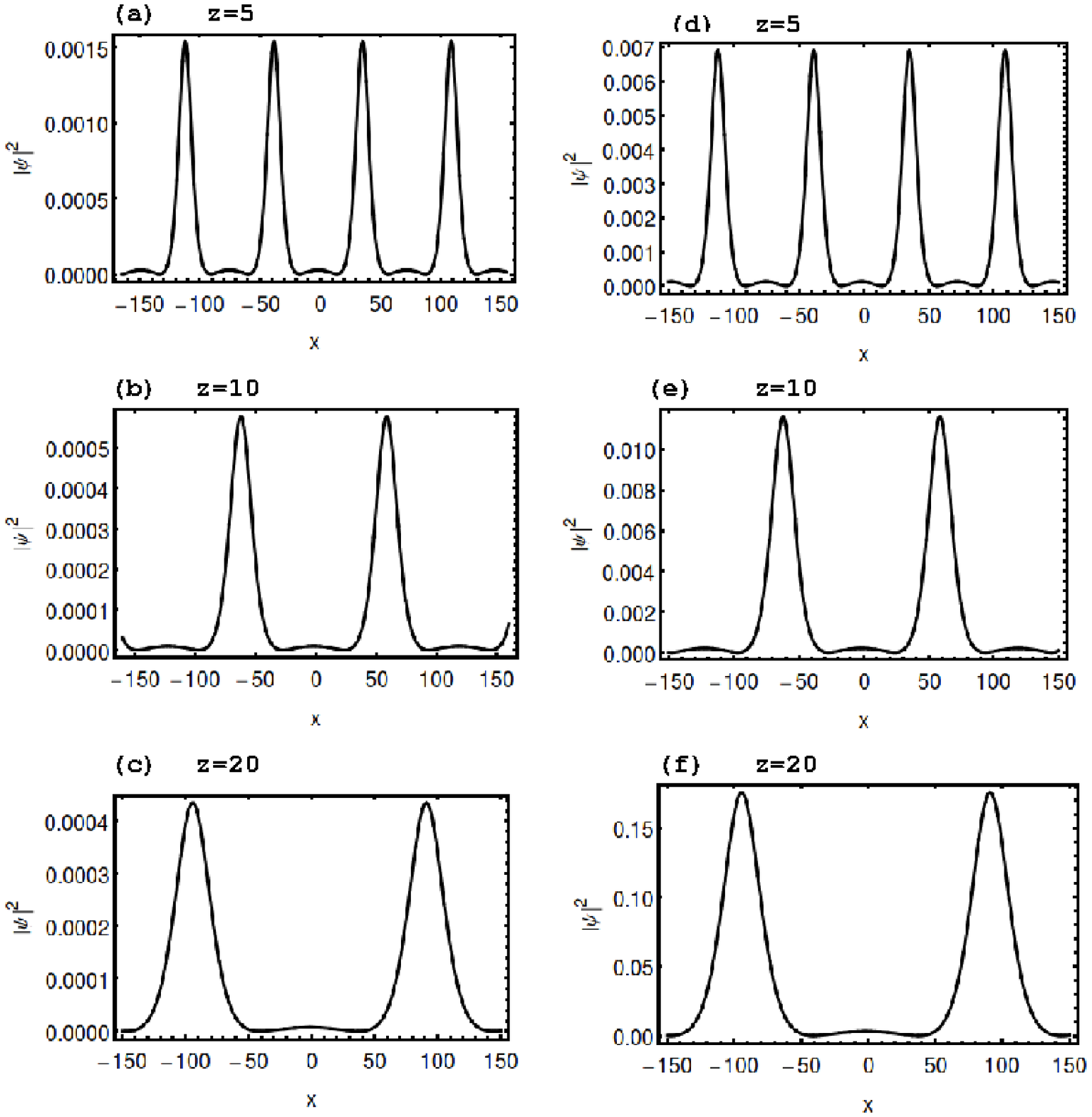}
\end{center}
\caption{The intensity profiles of AB for linear dispersion profile at (a) $z=5$, (b) $z=10$ and (c) $z=20$ with $\gamma_0=0.05$ and (d) $z=5$, (e) $z=10$, (f) $z=20$ with $\gamma_0=0.2$. The other parameters are fixed as same in Fig. \ref{3d:fig12}.}
\label{3d:fig13}
\end{figure*}
\begin{figure*}[!ht]
\begin{center}
\includegraphics[width=0.9\linewidth]{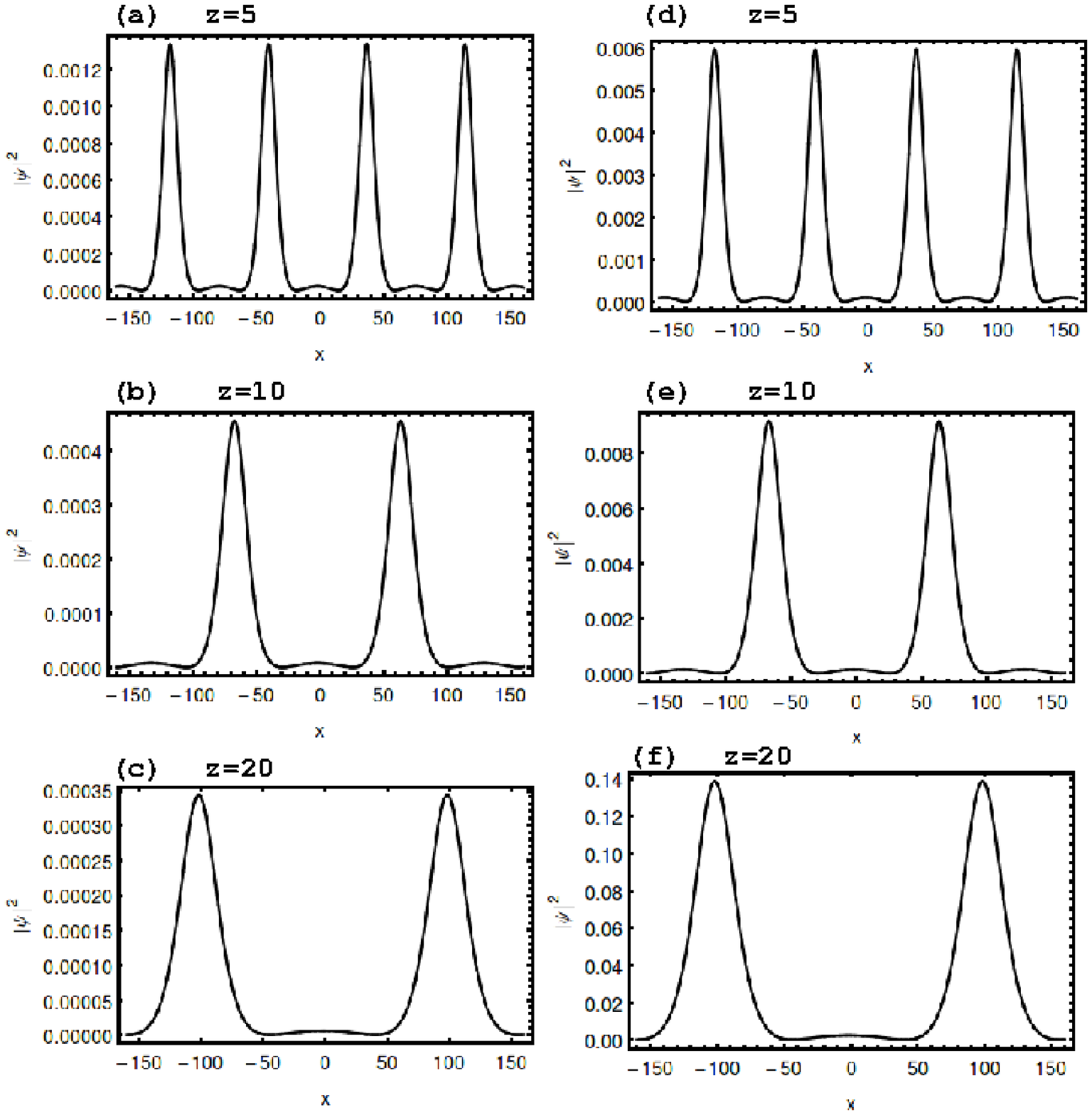}
\end{center}
\caption{The intensity profiles of AB for Gaussian dispersion profile at (a) $z=5$, (b) $z=10$ and (c) $z=20$ with $\gamma_0=0.05$ and (d) $z=5$, (e) $z=10$ and (f) $z=20$ with $\gamma_0=0.2$. The other parameters are fixed as same in Fig. \ref{3d:fig12}.}
\label{3d:fig14}
\end{figure*}

Fig. \ref{3d:fig12} shows the intensity profiles of periodic localized waves for the constant dispersion parameter at different propagation distances described by (\ref{3d:eq6}). The intensity of AB at three different distances, say $z=5$ (Fig. \ref{3d:fig12}(a)), $z=10$ (Fig. \ref{3d:fig12}(b)) and $z=20$ (Fig. \ref{3d:fig12}(c)) are depicted in this figure. At $z=5$, intensity of this periodic localized structure attains the value of maxima $|\psi|^2\approx0.025$ which is shown in Fig. \ref{3d:fig12}(a).  The intensity of AB and the number of peaks decreases at $z=10$ and $z=20$ as shown in Figs. \ref{3d:fig12}(b) and \ref{3d:fig12}(c).  When we tune the value of $\gamma_0$ to $0.2$ we observe that the intensity of this localized structure increases over the propagation distances, for example, at $z=5$ ($|\psi|^2\approx 0.12$), $z=10$ ($|\psi|^2\approx 0.27$) and $z=20$ ($|\psi|^2\approx 3.9$), as shown in Figs. \ref{3d:fig12}(d), \ref{3d:fig12}(e) and \ref{3d:fig12}(f).

\begin{table}[]
\centering
\caption{The effect of different dispersion profiles in the intensity of AB.}
\label{int-tab2}
\begin{tabular}{|c|c|c|}
\hline
\multirow{2}{*}{} & \multirow{2}{*}{} & Intensity \\ 
S.No. & Dispersion profile  &  of AB  \\ \hline
1 & constant & decreases   \\ \hline
2 & linear & decreases   \\ \hline
3 & Gaussian & decreases   \\ \hline
4 & hyperbolic & increases   \\ \hline
5 & logarithm & increases  \\ \hline
6 & exponent & increases \\ \hline 
\end{tabular}
\end{table}
 
\begin{figure*}[!ht]
\begin{center}
\includegraphics[width=0.9\linewidth]{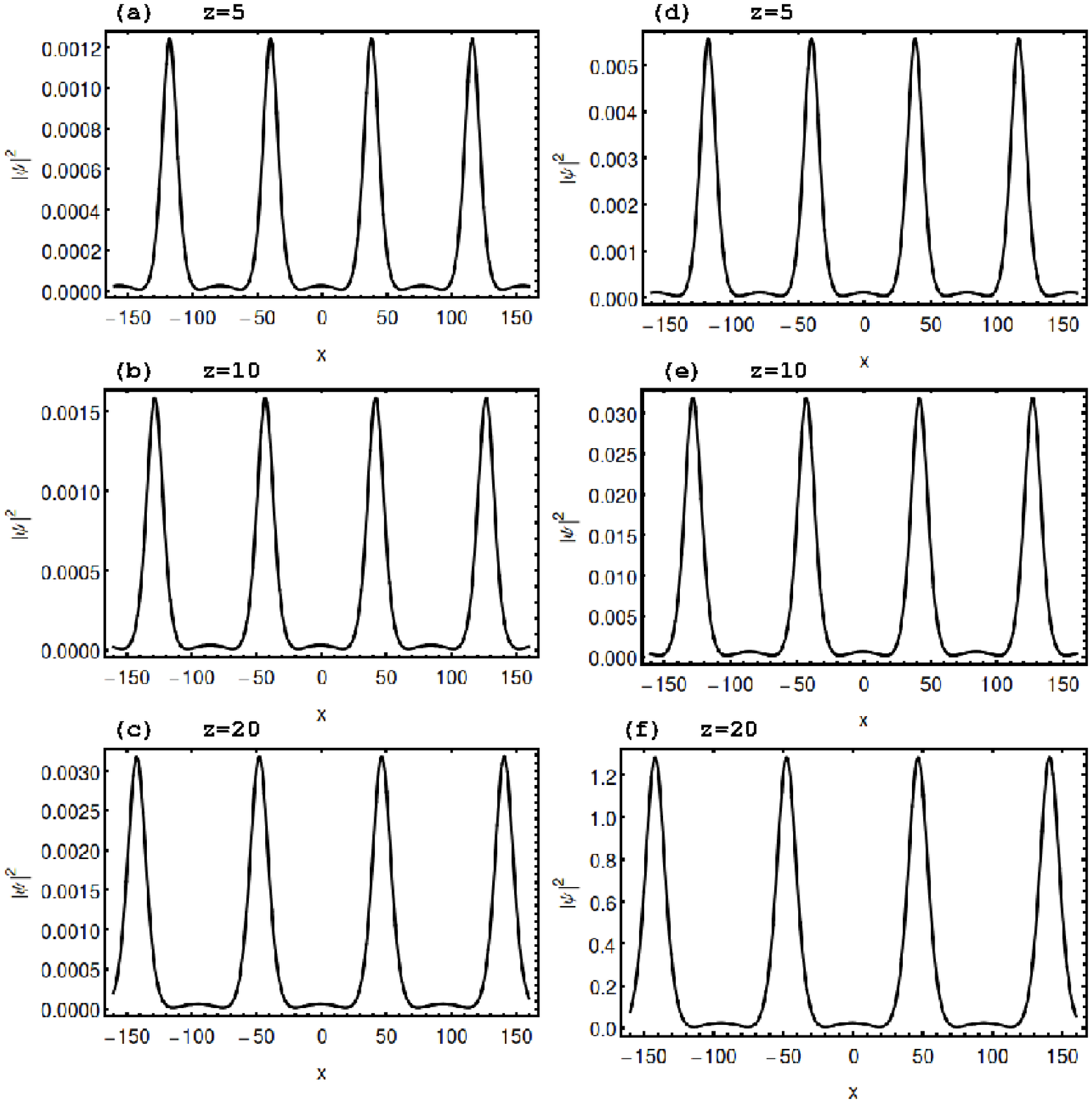}
\end{center}
\caption{The intensity profiles of AB for hyperbolic dispersion profile at (a) $z=5$, (b) $z=10$ and (c) $z=20$ with $\gamma_0=0.05$ and (d) $z=5$, (e) $z=10$ and (f) $z=20$ with $\gamma_0=0.2$. The other parameters are fixed as same in Fig. \ref{3d:fig12}.}
\label{3d:fig15}
\end{figure*}

\begin{figure*}[!ht]
\begin{center}
\includegraphics[width=0.9\linewidth]{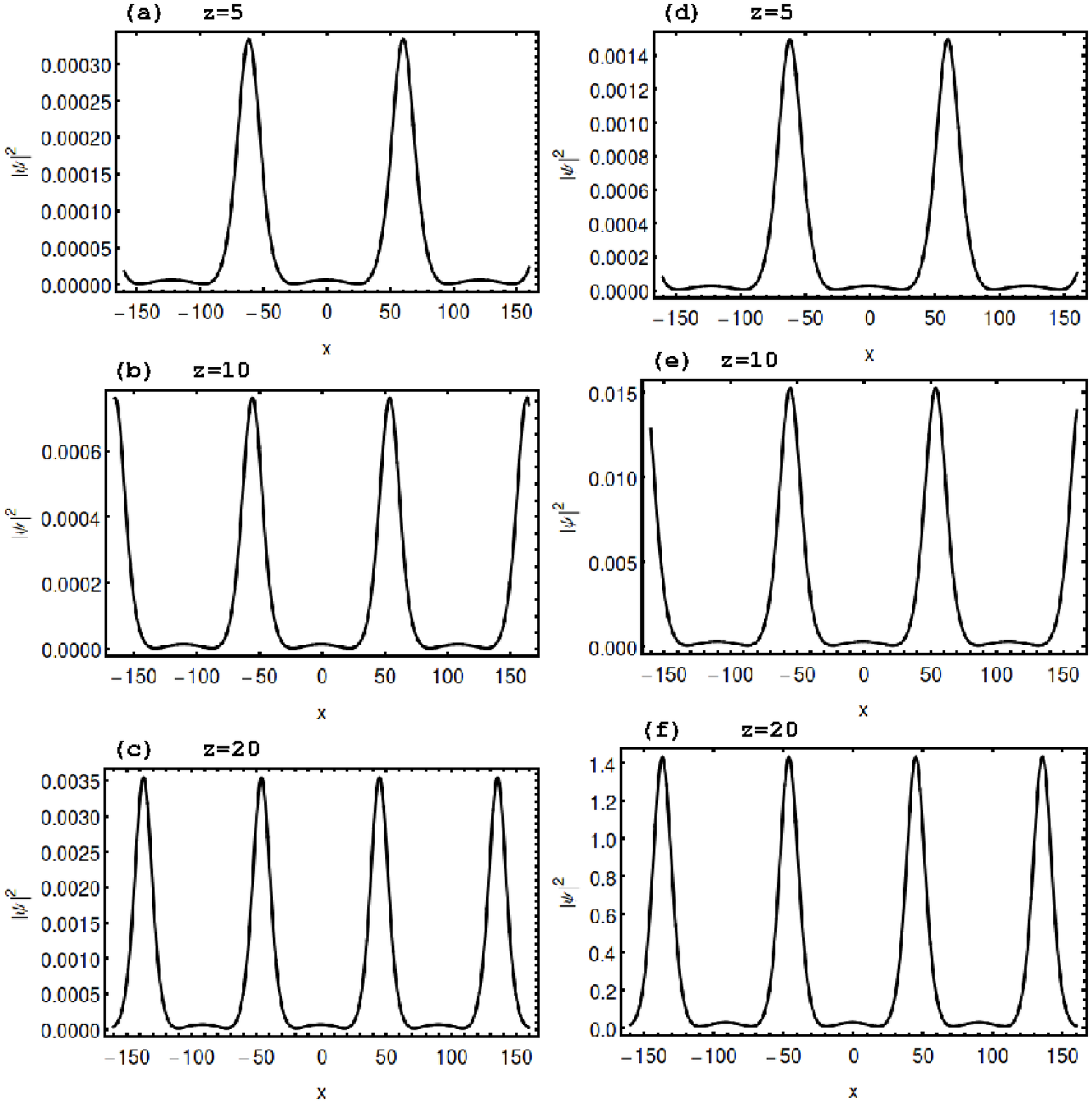}
\end{center}
\caption{The intensity profiles of AB for logarithmic dispersion profile at (a) $z=5$, (b) $z=10$ and (c) $z=20$ with $\gamma_0=0.05$ and (d) $z=5$, (e) $z=10$ and (f) $z=20$ with $\gamma_0=0.2$. The other parameters are fixed as same in Fig. \ref{3d:fig12}.}
\label{3d:fig16}
\end{figure*}
\begin{figure*}[!ht]
\begin{center}
\includegraphics[width=0.9\linewidth]{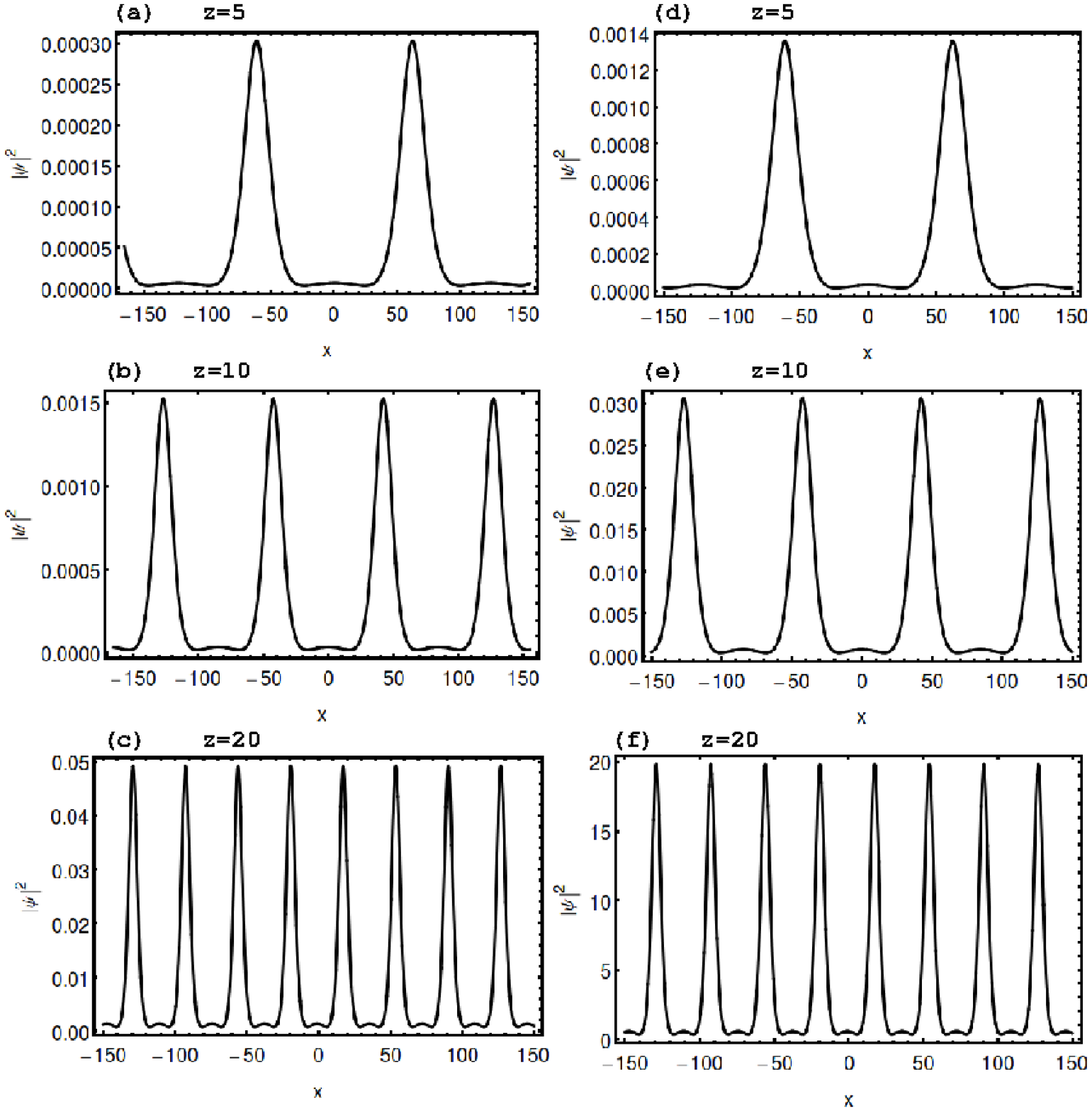}
\end{center}
\caption{The intensity profiles of AB for exponential dispersion profile at (a) $z=5$, (b) $z=10$ and (c) $z=20$ with $\gamma_0=0.05$ and (d) $z=5$, (e) $z=10$ and (f) $z=20$ with $\gamma_0=0.2$. The other parameters are fixed as same in Fig. \ref{3d:fig12}.}
\label{3d:fig17}
\end{figure*}
The propagation of AB through linear dispersion profile at three different distances, say $z=5$ (Fig. \ref{3d:fig13}(a)), $z=10$ (Fig. \ref{3d:fig13}(b)) and $20$ (Fig. \ref{3d:fig13}(c)), are shown in Fig. \ref{3d:fig13}.  Intensity of the above localized structure increases when we increase the value of $\gamma_0$ to $0.2$ as demonstrated in Figs. \ref{3d:fig13}(d), (e) and (f).  In Fig. \ref{3d:fig14}, we display the variation in the amplitude of AB for Gaussian type dispersion profile at three different propagation distances, that is $z=5, 10$ and $20$.  The intensity profile of AB at $z=5$ is shown in Fig. \ref{3d:fig14}(a).  The intensity of localized structure decreases at $z=10$ and $z=20$ but the number of peaks has reduced to $2$ as shown in Figs. \ref{3d:fig14}(b) and (c). When we increase the value of $\gamma_0$ to $0.2$ we observe that the intensity of AB increases as demonstrated in Figs. \ref{3d:fig14}(d), (e) and (f), respectively.

The profile of AB for hyperbolic dispersion parameter at three different distances, $z=5$ (Fig. \ref{3d:fig15}(a)), $z=10$ (Fig. \ref{3d:fig15}(b)) and $z=20$ (Fig. \ref{3d:fig15}(c)), are displayed in Fig. \ref{3d:fig15}.  We observe that the intensity of periodic localized structure increases with increasing propagation distances.  This is in contrast with the constant, linear and Gaussian dispersion profiles. When we increase the value of $\gamma_0$ from $0.05$ to $0.2$ we observe that the intensity of AB increases and the number of peaks remains the same in the considered region as seen in Figs. \ref{3d:fig15}(d), (e) and (f).  The variations in the amplitude of breather through logarithmic dispersion profile at three different distances, $z=5, 10$ and $20$, are presented in Fig. \ref{3d:fig16}.  The intensity of the ABs increases at $z=10 (0.0007)$ and $z=20 (0.0035)$ and the number of peaks increases in the considered region as shown in \ref{3d:fig16}(b) and (c).  When we tune the value of $\gamma_0$ to $0.2$, the intensity of periodic localized wave increases as seen in Figs. \ref{3d:fig16}(d), (e) and (f). 

The AB propagates through an exponential dispersion parameter at different distances, $z=5, 10$ and $20$ are displayed in Fig. \ref{3d:fig17}(a), (b) and (c). We note here that the intensity and the number of peaks increases over the propagation distances. When we tune the value of $\gamma_0$ to $0.2$, the intensity of AB increases as seen in Figs. \ref{3d:fig17}(d), (e) and (f). 

From the outcome, we conclude that the same initial pulse when it propagates through six different forms of dispersion parameters has attained different amplitudes and widths.  We found that the intensity of AB in the exponential dispersion profile turns out to be the largest one whereas its intensity in the Gaussian dispersion profile becomes the smallest.  The number of peaks in the periodic localized profile decreases in the constant, linear and Gaussian dispersion profiles and it increases/maintains in the hyperbolic, logarithmic and exponential dispersion profiles. Variations in the intensity of AB are tabulated in Table \ref{int-tab2} for the considered forms of $\beta(z)$. 
\section{Conclusion}
In this paper, we have constructed a family of rational solutions for the (3+1)-dimensional vcNLS equation with distributed coefficients such as  diffraction, nonlinearity and gain/loss parameter through similarity transformation.  The constructed rational solutions should satisfy a constraint in order to be the solution of three-dimensional vcNLS equation.  We have considered RW and breather solutions of NLS equation and investigated in detail the characteristics of these two profiles for six different forms of dispersion profiles.  Our results reveal that the intensity of the localized envelope solution attenuates over propagation distances in constant, linear, Gaussian dispersion profiles whereas it increases in hyperbolic and logarithmic dispersion profiles. For the case of exponential profile, the intensity of these localized envelope solutions becomes larger whereas it becomes smaller for logarithmic dispersion profile.  Our observations also reveal that the intensity of these localized envelope solutions increases when we increase the value of gain parameter.   We have also analyzed the trajectory of RW in each one of the cases separately.  For this purpose, we have obtained the expression for hump and valleys of the first-order localized envelope solution.  Our results reveal that the trajectories of the two valleys are not symmetric with respect to propagation distance in all the six profiles of $\beta(z)$.  In addition to the above, we have investigated the variation in the intensity of AB profile with six different forms of dispersion profiles. We found that the number of peaks in the periodic localized structure decreases in the constant, linear and Gaussian dispersion profiles whereas it increases/maintains in the hyperbolic, logarithmic and exponential dispersion profiles.  We remark that the solutions reported in this paper is a restricted class of solutions.  The solutions other than these, form singularities in finite time or blow-up. The obtained results may present novel properties to better understand corresponding localized wave phenomena in related fields.
\section*{Acknowledgements}
KM thanks the University Grants Commission (UGC-RFSMS), Government of India, for providing a research fellowship. The work of MS forms part of a research project sponsored by NBHM, Government of India.

\end{document}